\documentclass[10pt,a4paper]{article}
\usepackage[latin1, utf8]{inputenc}
\usepackage{amsmath}
\usepackage{setspace}
\usepackage{enumerate}
\usepackage[usenames,dvipsnames]{xcolor}
\usepackage{amsfonts}
\usepackage{amssymb}
\renewcommand{\bar}{\overline}
\usepackage{axodraw4j}
\usepackage{ltablex}
\usepackage[bottom]{footmisc}
\newcommand{\tensor}{\otimes}
\newcommand{\sym}{\textrm{Sym}}
\usepackage[section]{placeins}
\usepackage{pstricks}
\usepackage{mathtools}
\usepackage[labelfont=bf]{caption}
\usepackage{hyperref}
\newcommand{\dzero}{{\mathcal{O}(\delta^0)}}
\newcommand{\done}{{\mathcal{O}(\delta^1)}}
\usepackage{color}
\usepackage{cancel}
\newcommand{\K}{K}
\newcommand\numberthis{\addtocounter{equation}{1}\tag{\theequation}}
\newcommand{\A}{\mathcal{A}}
\usepackage{graphicx}
\usepackage{fullpage}
\usepackage{float}
\usepackage[nosort]{cite}

\newcommand{\ep}{\epsilon}
\usepackage{framed}

\renewcommand{\l}{\langle}
\renewcommand{\r}{\rangle}
\renewcommand{\a}{{\alpha}}
\newcommand{\ad}{{\dot\alpha}}

\newcommand{\tr}{\textrm{tree}}

\newcommand{\R}{\mathcal{R}}
\newcommand{\lp}{{1\textrm{-loop}}}
\newcommand{\mhv}{\textrm{MHV}}
\newcommand{\nmhv}{\textrm{NMHV}}
\usepackage{relsize}
\numberwithin{equation}{section}
\allowdisplaybreaks

\makeatletter
\g@addto@macro\bfseries{\boldmath}
\makeatother

\long\def\symbolfootnote[#1]#2{\begingroup%
\def\thefootnote{\fnsymbol{footnote}}\footnote[#1]{#2}\endgroup}

\def\beqa{\begin{eqnarray}}
\def\eeqa{\end{eqnarray}}
\def\beq{\begin{equation}}
\def\eeq{\end{equation}}

\hypersetup{
colorlinks=true,
linktoc=page,
citecolor=Blue,
linkcolor=Blue,
urlcolor=Blue} 

\begin{document}

% TITLE PAGE %

\begin{flushright}
QMUL-PH-15-21\\
\end{flushright}

\vspace{20pt}

\begin{center}

{\Large \bf One-Loop Soft Theorems via Dual Superconformal Symmetry  }\\
%\vspace{0.3 cm}
%{\Large \bf via Dual Superconformal Symmetry  }

\vspace{45pt}

{\mbox {\bf  Andreas Brandhuber,  Edward Hughes, Bill Spence and Gabriele Travaglini}}%
\symbolfootnote[4]{
{\tt  \{ \tt \!\!\!a.brandhuber, e.f.hughes, w.j.spence, g.travaglini\}@qmul.ac.uk}
}

\vspace{0.5cm}

\begin{center}
{\small \em

Centre for Research in String Theory\\
School of Physics and Astronomy\\
Queen Mary University of London\\
Mile End Road, London E1 4NS, UK
}
\end{center}

\vspace{40pt} 
{\bf Abstract}

\end{center}

\vspace{0.3cm} 

\noindent
We study soft theorems at one loop in planar $\mathcal{N}\!=\!4$ super Yang-Mills theory through finite order in the infrared regulator and to subleading order in the soft parameter $\delta$. In particular, we derive a universal constraint from dual superconformal symmetry, which we use to bootstrap subleading $\log \delta$ behaviour. Moreover, we determine the complete infrared-finite subleading soft  contribution  of $n$-point MHV amplitudes using momentum twistors. Finally, we compute the subleading $\log \delta$ behaviour of one-loop NMHV ratio functions at six and seven points, finding that  universality holds within but not between helicity sectors.
 
\setcounter{page}{0}
\thispagestyle{empty}
\newpage

% TABLE OF CONTENTS %

{\setstretch{1.3}
\setcounter{tocdepth}{4}
\hrule height 0.75pt
\tableofcontents
\vspace{0.8cm}
\hrule height 0.75pt
\vspace{1cm}

\setcounter{tocdepth}{2}
}

\section{Introduction}
Scattering amplitudes of massless particles in quantum field theory   have  infrared divergences, which are known to cancel in inclusive cross-sections   \cite{Bloch:1937pw,Kinoshita:1962ur, Lee:1964is}.
Exhibiting such cancellations in perturbation theory  relies on universal factorisation properties of amplitudes in the limit as the momentum of a massless particle is taken soft, $p\to \delta p$, which  makes the study of these limits particularly significant. 
\\\\
The leading soft behaviour was first studied at tree level in QED  \cite{Yennie:1961ad} and then in gravity  \cite{Weinberg:1965nx}.  Yang-Mills theories were also shown to exhibit similar factorisation \cite{Berends:1987me}. There is  however an important difference --  in gravity the leading soft behaviour  is not renormalised by loop effects \cite{Bern:1998sv}, unlike in QED and Yang-Mills. 
Parallel work considered the subleading soft corrections at tree level in QED and found universal behaviour described by a differential operator \cite{Low:1958sn, Burnett:1967km}. The analogous subleading correction in gravity was studied only recently using eikonal methods \cite{1103.2981}. 
Subsequently, in  \cite{1404.4091}  a simple universal form for the subleading and sub-subleading contributions for soft gravitons at tree level was discovered, which easily generalises to Yang-Mills theories \cite{1404.5551}.
\\\\
At one loop,  the leading soft behaviour is fully understood. Results  in QED  were found in \cite{Grammer:1973db} while the one-loop leading soft correction in QCD was  computed in  \cite{hep-ph/9810409,hep-ph/9903516} as part of a phenomenological study of NNLO jet production at colliders. 
Less is known about the subleading soft theorems at one-loop level. 
 This question was  first approached  in QED in  \cite{DelDuca:1990gz}. In QCD and gravity the infrared-divergent corrections were computed in   \cite{1405.1015} and found to be universal. 
A recent generalisation of \cite{DelDuca:1990gz}  hints at universality in the subleading $\log \delta $ terms of  gauge theories  \cite{1503.05156}. 
\\\\
Only partial information is available for the infrared-finite terms, both  in gauge theory and in  gravity. The subleading soft behaviour of certain rational amplitudes was computed in \cite{1405.1410}. Soft-collinear effective theory provides a route to a subleading theorem for soft gluon emission from well-separated hard particles in QCD \cite{1412.3108}. It is possible to evaluate the factorising contributions \cite{hep-ph/9503236} for soft graviton emission using locality and symmetry principles \cite{Broedel:2014bza}. Very recently the subleading soft behaviour of non-factorising terms was determined for single-minus graviton amplitudes through eight points \cite{1507.08882}. In this paper we evaluate the first complete $n$-point one-loop correction to subleading soft behaviour in a non-abelian gauge theory, namely planar $\mathcal{N}=4$ super Yang-Mills (SYM). 
The expressions we find, equations \eqref{111}--\eqref{222},  
contain
simple sums of bulk terms together with boundary contributions involving
legs near the soft particle. The latter give rise to a particularly simple
and universal form of the $\log \delta $ terms in the soft expansion. 
\\\\
Our modern understanding of soft theorems is  entwined with symmetry.  Indeed, the  soft behaviour was proved in   \cite{Strominger:2013jfa, He:2014laa} to be a Ward identity for BMS symmetry \cite{Bondi:1962px}, and it was further   conjectured  that the tree-level subleading soft graviton theorem emerges from a hidden Virasoro symmetry at null infinity. 
An interesting approach was pursued in 
 \cite{1405.2346}, where it was shown that  conformal symmetry is sufficient to determine the tree-level Yang-Mills subleading soft theorem. In fact, merely gauge and Poincar\'{e} invariance is enough \cite{1406.6987,1406.6574}. 
One of the goals of  this paper is to work out  the constraints imposed by dual conformal symmetry on soft theorems in $\mathcal{N}=4$ SYM at  tree level and one loop. 
This symmetry provides differential constraints on soft corrections, and
we have found these to be a powerful tool to determine them. Invoking some
straightforward and reasonable conjectures on the general form of the soft
corrections, the dual conformal symmetry requirements  are solvable and
the resulting expressions pass nontrivial tests.
The known simple form of the one-loop anomaly for dual superconformal
symmetry is a key element in this process -- this is a feature unavailable
for  conventional conformal symmetry.
\\\\
Several fascinating related papers lie slightly outside the main line of our development.  Various holographic theories now exist which manifest the tree-level soft theorems as Ward identities \cite{1405.5122, 1406.1462, 1407.3814, 1504.01364,  1503.02663, 1504.02304}. Subleading double-soft corrections have been studied in \cite{1503.04816, 1504.05558, 1504.05559, 1505.04411, 1507.00938}. Subleading soft behaviour has also been scrutinised in string theories \cite{1406.4172, 1406.5155, 1411.6661, 1502.05258, 1505.05854, 1505.08130}, and recently, subleading soft theorems have been extended to off-shell quantities as well \cite{1506.07551}.
\\\\
The rest of the paper is organised as follows. In Section 2 we review the  dual superconformal symmetry of $\mathcal{N}=4$ SYM. In Section 3 we summarise the supersymmetric soft gluon theorems at tree level and conjecture a form for a one-loop extension. In Section 4 we derive constraints on supersoft theorems at tree level and one loop using the anomalous dual conformal Ward identity for amplitudes. In Section 5 we  compute the subleading soft behaviour of general MHV amplitudes and the six- and seven-point NMHV amplitudes at one loop, employing  unitarity  and momentum twistors. 
In particular we present some evidence for universality of the subleading $\log \delta$ terms. We present our conclusions in Section 6 as well as suggestions  for future work. Several appendices are included which illustrate some of the technical points, and in particular Appendix \ref{package} documents a new Mathematica package used in Section~5.

\section{Dual Superconformal Symmetry}\label{dual-superconformal-symmetry}
In this section we recall some properties of the dual superconformal symmetry of amplitudes that we will need for later calculations. It is well known that planar colour-ordered gluon amplitudes in $\mathcal{N}=4$ SYM theory may equivalently be calculated as the expectation values of certain lightlike Wilson loops with appropriate operator insertions \cite{0705.0303, 0707.0243, 0707.1153, 1009.2225, 1010.1167}. Superconformal symmetry acting on the vertices of Wilson loops then yields a hidden dual superconformal symmetry of amplitudes \cite{0807.1095}.\\\\
We use on-shell superamplitudes \cite{Nair:1988bq} to make the symmetry manifest and arrange the external states into $\mathcal{N}=4$ supermultiplets, defining for particle $i$,
\begin{equation}
\begin{aligned}\label{superamp}
\Phi_i(p_i, \eta_i)&=G^+(p_i)+\eta_i^A\lambda_A(p_i)+\frac{1}{2}\eta_i^A\eta_i^B S_{AB}(p_i)\\
&+\frac{1}{3!}\eta_i^A\eta_i^B\eta_i^C\epsilon_{ABCD}\bar\lambda^D(p_i)+\frac{1}{4!}\eta_i^A\eta_i^B\eta_i^C\eta_i^D \epsilon_{ABCD}G^-(p_i)\, ,
\end{aligned}
\end{equation}
where $\eta_i^A$ are auxiliary Grassmann variables and $A=1, \ldots , 4$ is an $SU(4)$ index. A superamplitude in on-shell coordinates $(|i],|i\r, \eta_i)$ is then 
\begin{equation}
\mathcal{A}_n(|i],|i\r, \eta_i)=\mathcal{A}(\Phi_1,\dots, \Phi_n) \, .
\end{equation} 
To exhibit the dual superconformal symmetry, we introduce dual coordinates $(x_i, \theta_i)$ \cite{0807.1095} defined by 
\begin{equation}
(x_i - x_{i+1})^{\ad\alpha} = | i \r^\alpha [i|^\ad  \,  , \quad \qquad (\theta_i-\theta_{i+1})^{\alpha A}  = | i \r^\alpha \eta_i^A \, .
\end{equation}
Momentum conservation and supersymmetry imply that $(x_{n+1},\theta_{n+1})=(x_1,\theta_1)$ for an $n$-particle scattering process. Assuming that the amplitude is written exclusively in terms of the coordinates $(|i\r, |i],\eta_i)$, the generator of dual conformal boosts takes the form,
\begin{equation}\label{dual-conformal-boost-generator}
K_{\alpha\dot\alpha} = \sum_{i=1}^n\Big( x_
{i\dot\alpha}^\beta\l i|_\alpha\frac{\partial}{\partial|i\r^\beta}+x_{i+1
\alpha\dot\beta}|i]_\ad \frac{\partial}{\partial|i]_{\dot\beta}}+\theta_{i+1 \alpha}^A|i]_\ad\frac{\partial}{\partial \eta_i^A}\Big)\, .
\end{equation}
Tree amplitudes transform covariantly as
\begin{equation}\label{dual-conformal-covariance}
K_{\alpha\ad}\A_n^\tr=-\Big(\sum_{i=1}^n x_{i\alpha\ad}\Big)\A_n^\tr\, .
\end{equation}
Note that the amplitude contains the universal factor $\delta^{(4)}(p)\delta^{(8)}(q)$ which is dual conformal invariant. 
\\\\
At loop level the symmetry is anomalous, owing to divergences in the Wilson loop \cite{Polyakov:1980ca}. In particular the infrared divergences of loop amplitudes correspond  to the ultraviolet cusp divergences of  Wilson loops \cite{Korchemsky:1985xj}. This provides us with a useful means to visualise the entanglement between subleading soft behaviour and infrared divergences at loop level, as we will discuss in Section \ref{1-lp-soft}. 
\\\\
The anomalous Ward identities calculated in \cite{0709.2368,0712.1223} suffice to explain the  BDS ansatz for all-loop MHV amplitudes \cite{hep-th/0505205}, which is correct up to a function of dual conformal invariant cross-ratios. We will require the explicit form of the one-loop anomaly proved in \cite{0906.3552}, namely
\begin{equation}\label{dual-conformal-anomaly}
K_{\alpha \ad} \A_n^\lp \ =  \ \frac{2}{\epsilon}c_\Gamma 
\A_n^\tr \sum_{i=1}^n x_{i\alpha\ad} \big[-\vphantom{p_i^j}(i-1 \ i )\big]^{-\epsilon} \, - \, \A_n^\lp \sum_{i=1}^n x_{i\alpha \ad} \, ,
\end{equation}
valid through $\mathcal{O}(\epsilon^0)$, where $(i \ j) := 2 p_i \cdot p_j$, $\epsilon$ is an infrared regulator,  and 
\begin{equation}
c_\Gamma=\frac{\Gamma(1+\epsilon)\Gamma^2(1-\epsilon)}{(4 \pi)^{2-\epsilon}\Gamma(1-2\epsilon)}\, .
\end{equation}

\section{Summary of Soft Gluon Theorems}\label{supersoft-thms}
In preparation for the new results presented in Sections 4 and 5, here we  recall some known results  on  soft limits of superamplitudes.

\subsection{Tree Level}\label{tree-soft-sec}
Consider the holomorphic soft limit of a positive-helicity gluon $n^+$ in an $n$-particle amplitude, 
\begin{equation}
|n\r \to \delta |n\r \ , \quad |n] \to |n] \ , \quad p_n \to \delta p_n\, ,
\end{equation}
where%
\footnote{See Appendix \ref{spinor-conventions} for our spinor helicity conventions.}
 $p_n^{\alpha\ad} = |i\r^\alpha [i|^{\ad}$.   
Let $\mathcal{A}_n$ denote an $n$-particle colour-ordered   superamplitude in planar $\mathcal{N}=4$ SYM. Then expanding in $\delta$  one has, at tree level \cite{1404.5551}, 
\begin{equation}\label{tree-soft-theorem-cs}
\A_n^\tr \to \left(\frac{1}{\delta^2}S^{(0)}+\frac{1}{\delta}S^{(1)}\right)\A_{n-1}^\tr \, ,
\end{equation}
where $S^{(0)}$ and $S^{(1)}$ are given by 
\begin{align}
S^{(0)}&=\frac{\l n-1 \ 1\r}{\l n-1 \ n \r\l n \ 1 \r}\label{soft-op-leading}\, ,\\
S^{(1)}&=\frac{|n]}{\l n-1 \ n \r}\cdot\frac{\partial}{\partial |n-1]}+\frac{|n]}{\l n \ 1 \r}\cdot\frac{\partial}{\partial |1]}%+\frac{\eta_n^A}{\l n-1 \ n \r}\frac{\partial}{\partial\eta_{n-1}^A}+\frac{\eta_n^A}{\l n \ 1 \r}\frac{\partial}{\partial\eta_1^A}
\, .
\label{soft-op-subleading}
\end{align}
Note  that these operators are antisymmetric about particle $n$. This is a consequence of our freedom to relabel particles in the opposite direction without changing the physics.%
\footnote{Soft theorems can also be formulated without restricting leg $n$ to be a gluon. Doing this one finds that for scalar particles there is no soft theorem, while for gluinos there is only a leading-order ($1 / \delta$) statement \cite{1410.1616}. } 
\\\\
To perform practical calculations it is convenient to work with stripped amplitudes $A_n$, where
\begin{equation}
\A_n  =  A_n \, \delta^{(4)}(P_n) \, ,
\end{equation}
with  $P_n:= \sum_{i=1}^n p_i$. Note that a stripped amplitude is ambiguous without a momentum conservation prescription. One means of resolving this is by eliminating two antiholomorphic spinors $|a]$ and $|b]$ \cite{1404.4091}. We may define such an elimination for any function $f$ of external kinematics as
\begin{equation}\label{stripped-mom-cons}
f^{(ab),n}=\int\!\!d|a]d|b] \  \ |\l a \ b \r|  \,  \delta^{(4)}(P_n) \, f \, ,
\end{equation}
so that an unambiguous stripped amplitude may be written as $A_n^{(ab),n}$. Clearly it is useful to have an explicit prescription for performing the integral in (\ref{stripped-mom-cons}). We impose the equalities,
\begin{equation}\label{mom-cons-prescription}
|a] = \frac{1}{\l a \ b \r}\sum_{i\neq a}^n\l b \ i\r |i], \qquad |b] = \frac{1}{\l b \ a \r}\sum_{i\neq b}^n\l a \ i\r |i] \, .
\end{equation}
These relations are especially important when considering the soft behaviour  at one loop, which turns out to depend on the choice of $|a]$ and $|b]$.%
\footnote{In other words, it depends on how one implements momentum conservation, in a way similar to stripped amplitudes. }
\\\\ 
Taking the integrals through the derivatives in (\ref{tree-soft-theorem-cs}) proves the result for stripped amplitudes, 
\begin{equation}\label{tree-soft-theorem-cs-integrated}
A_n^{\tr(ab),n}\ \to  \ \left(\frac{1}{\delta^2}S^{(0)}+\frac{1}{\delta}S^{(1)}\right)A_{n-1}^{\tr(ab),n-1}\, ,
\end{equation}
as found  in \cite{1404.4091} in the case of gravity. 
\\\\
In \cite{1405.1015} Bern, Nohle and Davies argued for a statement equivalent to (\ref{tree-soft-theorem-cs}), namely
\begin{equation}\label{tree-soft-theorem-bnd}
\A_n^{\tr}\ \to\  \delta^{(4)}(P_n)\left(\frac{1}{\delta^2}S^{(0)}+\frac{1}{\delta}S^{(1)}\right)A_{n-1}^\tr\, , 
\end{equation}
where the momentum conservation delta function sits in front of the soft operator. 
The distinguishing property of this expression is that it features $n$-point momentum conservation on both sides. Many explicit examples have been calculated in the literature demonstrating the equivalence of (\ref{tree-soft-theorem-cs}) and (\ref{tree-soft-theorem-bnd}) and the issue was discussed formally in \cite{1406.6574}. One may  verify this equivalence by Taylor expanding $\delta^{(4)}(P_n)$ and applying the chain rule to $S^{(1)}\delta^{(4)}(P_{n-1})$. \\\\
We may easily write down the stripped amplitude version of (\ref{tree-soft-theorem-bnd});  that is
\begin{equation}\label{tree-soft-theorem-bnd-integrated}
A_n^{\tr(ab),n}\to \left[\left(\frac{1}{\delta^2}S^{(0)}+\frac{1}{\delta}S^{(1)}\right)A_{n-1}^\tr\right]^{(ab),n} \, .
\end{equation}
This formulation has an advantage over (\ref{tree-soft-theorem-cs-integrated}) because it allows one to adopt the following two step strategy to verify soft theorems:
\begin{enumerate}
\item Choose arbitrary forms for $A_n$ and $A_{n-1}$ and determine
\beq
\label{up}
A_n- {1\over  \delta^2}S^{(0)}A_{n-1}-{1 \over  \delta} S^{(1)}A_{n-1}\  . 
\eeq
\item Apply $n$-point momentum conservation and expand in $\delta$, 
then one  finds zero up to $\mathcal{O}(\delta^0)$.
\end{enumerate}
We emphasise that this approach leads to so-called feed-down terms from Taylor-expanding the term 
\begin{equation}\label{feed-down}
\left[-\frac{1}{\delta^2}S^{(0)}A_{n-1}^{\rm tree} \right]^{(ab),n}\, ,
\end{equation}
in \eqref{up} evaluated using $n$-point momentum conservation (which contains $\delta$-dependence).  
In this paper we shall consider soft theorems only in the language of (\ref{tree-soft-theorem-bnd}) and (\ref{tree-soft-theorem-bnd-integrated}), which is better suited to a loop-level generalisation.

\subsection{One Loop}\label{1-lp-soft}
At one-loop level, the leading soft behaviour is well-known \cite{hep-ph/9810409,hep-ph/9903516,hep-th/0510253}. Subleading soft theorems for the infrared-divergent part of generic one-loop amplitudes were found in \cite{1405.1015}. Based on this, one may conjecture the one-loop extension to the subleading soft theorem,
\begin{equation}\label{loopsoftthm}
A_n^\lp \to  \frac{1}{\delta^2}\left(S^{(0)} A_{n-1}^\lp + S^{(0)\lp} A_{n-1}^\tr\right)+\frac{1}{\delta}\left(S^{(1)}A_{n-1}^\lp + S^{(1)\lp}A_{n-1}^\tr\right) \, ,
\end{equation}
where the leading soft factor is  \cite{hep-ph/9810409,hep-ph/9903516}
\begin{equation}\label{S0lp}
S^{(0)\lp} \ = \ S^{(0)}F^{(0)}\ , 
\qquad F^{(0)} \ = \ \left(\frac{c_\Gamma}{\epsilon^2}\frac{\pi \epsilon}{\sin (\pi \epsilon)}\right)\left(-\frac{1}{\delta^2}\frac{(n-1 \ 1 )}{(n-1 \ n )(n \ 1)}\right)^\epsilon \, ,
\end{equation}
and the infrared-divergent part of the subleading soft operator is   \cite{1405.1015}

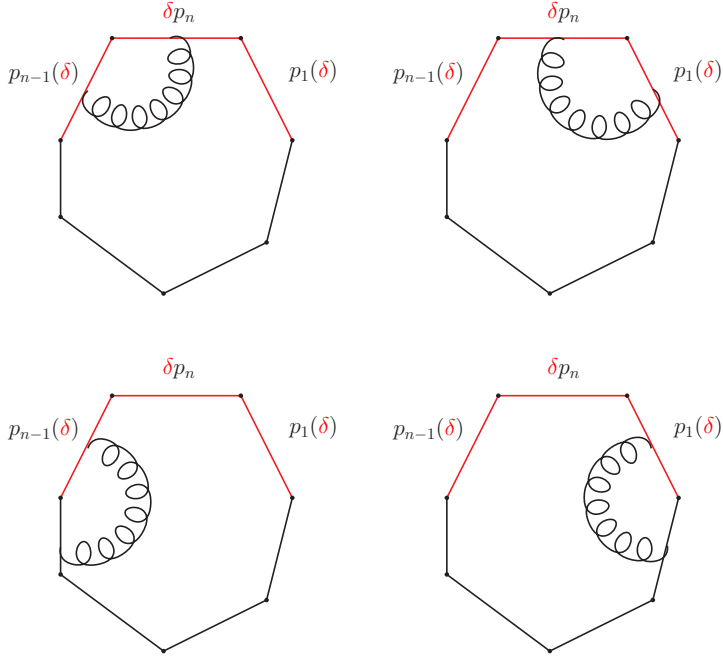
\begin{figure}[t]
%[!htb]
\begin{center}
\scalebox{0.6}{
\fcolorbox{white}{white}{
  \begin{picture}(447,418) (51,-48)
    \SetWidth{1.0}
    \SetColor{Red}
    \Line(80,274)(112,338)
    \Line(112,338)(192,338)
    \Line(192,338)(224,274)
    \SetColor{Black}
    \Line(224,274)(208,210)
    \Line(208,210)(144,178)
    \Line(144,178)(80,226)
    \Line(80,274)(80,226)
    \SetWidth{0.0}
    \Vertex(112,338){1.414}
    \Vertex(192,338){1.414}
    \Vertex(224,274){1.414}
    \Vertex(208,210){1.414}
    \Vertex(144,178){1.414}
    \Vertex(80,226){1.414}
    \Vertex(80,274){1.414}
    \Text(144,349)[lb]{\Large{\Black{${\color{red}\delta} p_n$}}}
    \Text(223,311)[lb]{\Large{\Black{$p_1({\color{red}\delta})$}}}
    \Text(48,310)[lb]{\Large{\Black{$p_{n-1}({\color{red}\delta})$}}}
    \SetWidth{1.0}
    \SetColor{Red}
    \Line(320,274)(352,338)
    \Line(352,338)(432,338)
    \Line(432,338)(464,274)
    \SetColor{Black}
    \Line(464,274)(448,210)
    \Line(448,210)(384,178)
    \Line(384,178)(320,226)
    \Line(320,274)(320,226)
    \SetWidth{0.0}
    \Vertex(352,338){1.414}
    \Vertex(432,338){1.414}
    \Vertex(464,274){1.414}
    \Vertex(448,210){1.414}
    \Vertex(384,178){1.414}
    \Vertex(320,226){1.414}
    \Vertex(320,274){1.414}
    \Text(384,349)[lb]{\Large{\Black{${\color{red}\delta} p_n$}}}
    \Text(463,311)[lb]{\Large{\Black{$p_1({\color{red}\delta})$}}}
    \Text(288,310)[lb]{\Large{\Black{$p_{n-1}({\color{red}\delta})$}}}
    \SetWidth{1.0}
    \SetColor{Red}
    \Line(80,50)(112,114)
    \Line(112,114)(192,114)
    \Line(192,114)(224,50)
    \SetColor{Black}
    \Line(224,50)(208,-14)
    \Line(208,-14)(144,-46)
    \Line(144,-46)(80,2)
    \Line(80,50)(80,2)
    \SetWidth{0.0}
    \Vertex(112,114){1.414}
    \Vertex(192,114){1.414}
    \Vertex(224,50){1.414}
    \Vertex(208,-14){1.414}
    \Vertex(144,-46){1.414}
    \Vertex(80,2){1.414}
    \Vertex(80,50){1.414}
    \Text(144,125)[lb]{\Large{\Black{${\color{red}\delta} p_n$}}}
    \Text(223,87)[lb]{\Large{\Black{$p_1({\color{red}\delta})$}}}
    \Text(48,86)[lb]{\Large{\Black{$p_{n-1}({\color{red}\delta})$}}}
    \SetWidth{1.0}
    \SetColor{Red}
    \Line(320,50)(352,114)
    \Line(352,114)(432,114)
    \Line(432,114)(464,50)
    \SetColor{Black}
    \Line(464,50)(448,-14)
    \Line(448,-14)(384,-46)
    \Line(384,-46)(320,2)
    \Line(320,50)(320,2)
    \SetWidth{0.0}
    \Vertex(352,114){1.414}
    \Vertex(432,114){1.414}
    \Vertex(464,50){1.414}
    \Vertex(448,-14){1.414}
    \Vertex(384,-46){1.414}
    \Vertex(320,2){1.414}
    \Vertex(320,50){1.414}
    \Text(384,125)[lb]{\Large{\Black{${\color{red}\delta} p_n$}}}
    \Text(463,87)[lb]{\Large{\Black{$p_1({\color{red}\delta})$}}}
    \Text(288,86)[lb]{\Large{\Black{$p_{n-1}({\color{red}\delta})$}}}
    \SetWidth{1.0}
    \GluonArc[clock](124.526,318.369)(30.601,39.905,-154.095){7.5}{7}
    \GluonArc[clock](95.587,48.057)(32.974,87.545,-118.211){7.5}{7}
    \GluonArc(416.529,314.454)(32.587,136.222,344.964){7.5}{7}
    \GluonArc(444.627,49.291)(31.797,85.719,292.9){7.5}{7}
  \end{picture}
}}
\end{center}
\caption{\it The four  diagrams contributing to the infrared-divergent terms in the soft theorem at one loop.}
\label{IR-divergent-Wilson-loop}
\end{figure}
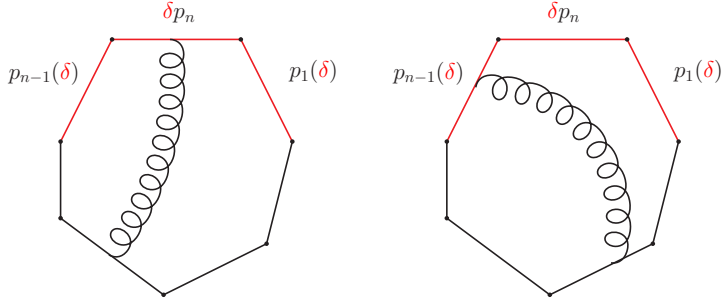
\begin{figure}[t]
%[!htb]
\begin{center}
\scalebox{0.6}{
\fcolorbox{white}{white}{
  \begin{picture}(447,194) (51,-48)
    \SetWidth{1.0}
    \SetColor{Red}
    \Line(80,50)(112,114)
    \Line(112,114)(192,114)
    \Line(192,114)(224,50)
    \SetColor{Black}
    \Line(224,50)(208,-14)
    \Line(208,-14)(144,-46)
    \Line(144,-46)(80,2)
    \Line(80,50)(80,2)
    \SetWidth{0.0}
    \Vertex(112,114){1.414}
    \Vertex(192,114){1.414}
    \Vertex(224,50){1.414}
    \Vertex(208,-14){1.414}
    \Vertex(144,-46){1.414}
    \Vertex(80,2){1.414}
    \Vertex(80,50){1.414}
    \Text(144,125)[lb]{\Large{\Black{${\color{red}\delta} p_n$}}}
    \Text(223,87)[lb]{\Large{\Black{$p_1({\color{red}\delta})$}}}
    \Text(48,86)[lb]{\Large{\Black{$p_{n-1}({\color{red}\delta})$}}}
    \SetWidth{1.0}
    \SetColor{Red}
    \Line(320,50)(352,114)
    \Line(352,114)(432,114)
    \Line(432,114)(464,50)
    \SetColor{Black}
    \Line(464,50)(448,-14)
    \Line(448,-14)(384,-46)
    \Line(384,-46)(320,2)
    \Line(320,50)(320,2)
    \SetWidth{0.0}
    \Vertex(352,114){1.414}
    \Vertex(432,114){1.414}
    \Vertex(464,50){1.414}
    \Vertex(448,-14){1.414}
    \Vertex(384,-46){1.414}
    \Vertex(320,2){1.414}
    \Vertex(320,50){1.414}
    \Text(384,125)[lb]{\Large{\Black{${\color{red}\delta} p_n$}}}
    \Text(463,87)[lb]{\Large{\Black{$p_1({\color{red}\delta})$}}}
    \Text(288,86)[lb]{\Large{\Black{$p_{n-1}({\color{red}\delta})$}}}
    \SetWidth{1.0}
    \GluonArc[clock](-26.904,90.384)(176.491,7.69,-39.132){7.5}{10}
    \GluonArc[clock](344.215,1.673)(82.561,94.317,-19.584){7.5}{10}
  \end{picture}
}}
\end{center}
\caption{\it Two of the $3n{-}10$ diagrams contributing to the finite terms in the soft theorem at one loop.}
\label{IR-finite-Wilson-loop}
\end{figure}


\begin{equation}\label{S1lp}
\begin{aligned}
\left.S^{(1)\lp}\right|_{\textrm{div.}}&=\frac{c_\Gamma}{\epsilon^2}\left[1+\epsilon\log\left(-\frac{1}{\delta^2}\frac{(n-1 \ 1 )}{(n-1 \ n )(n \ 1)}\right)\right]S^{(1)\tr}
\\
&+\frac{c_\Gamma}{\epsilon}\left[\frac{[n-1 \ n]}{[n-1 \ 1 ]\l 1 \ n \r}+\frac{[2 \ n ]}{[2 \ 1 ]\l 1 \ n \r}-\frac{[1 \ n]}{[1 \ n-1]\l n-1 \ n \r}-\frac{[n-2 \ n]}{[n-2 \ n-1]\l n-1 \ n \r}\right]\, .
\end{aligned}
\end{equation}
More generally we conjecture that the subleading soft operator takes the form,
\begin{equation}\label{subsoftZ}
S^{(1)\lp}\ =\ F^{(1)}S^{(1)}\, +\, c_\Gamma Z \, .
\end{equation}
Here $F^{(1)}$ and $Z$ are functions of external kinematics. We shall refer to $Z$ as the 
{\it subleading soft anomaly}.  
Note that $Z$ is only defined up to a momentum conservation prescription -- it is frame dependent. Nevertheless it remains a useful and practical quantity. Indeed we may immediately transform $Z$ between frames using of the elimination (\ref{mom-cons-prescription}). In Section \ref{main} we will fix $F^{(1)}$ and derive a differential constraint on $Z$. Section \ref{verification} then provides explicit computations of $Z$ for amplitudes in the MHV and NMHV sectors. All of our results will be valid through finite order in $\epsilon$.
\\\\
From a Wilson loop perspective, one-loop  amplitudes decompose into a sum of diagrams with one internal gluon. Evaluating each diagram requires a ultraviolet regulator $\epsilon$ which corresponds exactly to the infrared regulator of the loop amplitude. Only diagrams in which a gluon attaches to a $\delta$-dependent external momentum will contribute to the one-loop soft anomaly. 
\\\\
It is useful to distinguish the diagrams in which the internal gluon connects adjacent edges of the polygon. These have a ultraviolet cusp divergence, and in fact capture all infrared-divergent terms in the amplitude \cite{Korchemsky:1985xj}. This restriction  limits the number of diagrams required to analyse the infrared-divergent soft anomaly. In fact, choosing a symmetric momentum conservation prescription eliminating $(|n-1],|1])$ we see that the four  diagrams in Figure \ref{IR-divergent-Wilson-loop} suffice.
\\\\
The remaining diagrams generate the finite parts of box functions \cite{0707.1153, 1009.2225, 1010.1167}. Examples are displayed in Figure \ref{IR-finite-Wilson-loop}. It is important to note a conceptual subtlety: although the terms from these diagrams are independent of $\epsilon$ they still contribute to the subleading soft anomaly.  The large number of contributing diagrams makes finite order analysis significantly harder; nevertheless in Section \ref{verification} we shall see surprising cancellations leading to compact formulae. 

\section{Dual Superconformal Constraints on Soft Theorems}\label{main}
\subsection{Summary of Results}\label{main-summary}
In this section we derive dual superconformal constraint equations for leading and subleading soft operators at tree level and one loop. We collect the results here for simplicity; the notation used is defined subsequently where it is new.
\\\\
At tree level and leading  order in the soft parameter $\delta$,  we find that
\begin{equation}
\big( K_{\a \ad}\big)_{\mathcal{O}(\delta^0)}S^{(0)}=\Big(\sum_{j=3}^{n-1}|j]\l j| \Big)S^{(0)}\, ,
\end{equation}
while at subleading order in $\delta$,
\begin{equation}
\begin{aligned}
&-\frac{|n]\l 1 |}{\l n \ 1 \r}A_{n-1}^\tr+S^{(0)}\left(K_{\a \ad}\right)_{\mathcal{O}(\delta^1)}A_{n-1}^\tr+\left[\left(K_{\a \ad}\right)_{\mathcal{O}(\delta^0)},S^{(1)\tr}\right]A_{n-1}^\tr\\&+A_{n-1}^\tr S^{(1)\tr}\left(\sum_{i\neq 3,n}\sum_{j=3}^{i-1} {}^{'} |j]\l j|\right)=2|n]\l n| S^{(0)}A_{n-1}^\tr+\left(\sum_{j=3}^{n-1}|j]\l j|\right)S^{(1)\tr}A_{n-1}^\tr \, . 
\end{aligned}
\end{equation}
At one loop and leading order in $\delta$,
\begin{equation}
\left(K_{\a \ad}\right)_{\mathcal{O}(\delta^0)}S^{(0)\lp}=\frac{2}{\epsilon}c_\Gamma\left(\sum_{j=3}^{n-1} |j]\l j|\right)
\left[\vphantom{\frac{A}{B}}\left(-\vphantom{p_i^j}\delta(n-1 \ n)\right)^{-\epsilon}+\left(-\vphantom{p_i^j}\delta (n \ 1)\right)^{-\epsilon}-\left(-\vphantom{p_i^j}(n-1 \ 1 )\right)^{-\epsilon}\right] \, ,
\end{equation}
while at    subleading order in $\delta$,
\begin{equation}\label{cp}
\begin{aligned}
&S^{(0)}A_{n-1}^\tr\left(\K_{\a \ad}\right)_{\mathcal{O}(\delta^1)}F^{(0)}
+A_{n-1}^\tr\left(\K_{\a \ad}\right)_{\mathcal{O}(\delta^0)}Z =\left(\textrm{anomaly}\right)_{\mathcal{O}(\delta^1)}S^{(0)}A_{n-1}^\tr\\
&-S^{(1)\tr}\left[\left(\textrm{anomaly}\right)_{n-1}\right]A_{n-1}^\tr
 +\left[\left(\textrm{covariance}\right)_\dzero-\left(\textrm{covariance}\right)_{n-1} \right]ZA_{n-1}^\tr \, .
 \end{aligned}
\end{equation}
We verify that known expressions satisfy our formulae and conversely argue that the constraints determine ans\"atze for the operators. Finally, we propose the form of the $\log \delta $ part of the unknown infrared-finite soft anomaly,
\begin{equation}
Z_0|_{\log\delta}= \left(\frac{( n \ 1 )}{(n-1 \ 1)}+\frac{(n-2 \ n)}{(n-2 \ n-1 )}-\frac{(n- 2 \ 1)(n-1 \ n )}{(n-2 \ n-1)(n-1 \ 1)}\right)S^{(0)}\log(-(n-1 \ n)) + (i \leftrightarrow n-i) \, .
\end{equation}

\subsection{Soft Theorems from Conformal Symmetry}
In \cite{1405.2346},   conformal symmetry was used in order to determine the tree-level soft theorem (\ref{tree-soft-theorem-bnd}). As a warm-up to our dual conformal calculations we shall briefly review this method. From now on  we employ arbitrary forms of the stripped amplitudes, with the proviso that an $n$-point momentum conservation prescription should be applied afterwards. 
\\\\
First recall that the conformal generator takes the form\footnote{Here, and elsewhere in this section, we leave some spinor indices implicit.},
\begin{equation}\label{lark-start}
k_{\alpha \ad}=\sum_{i=1}^n\frac{\partial^2}{\partial |i\r\partial [i|} \, ,
\end{equation}
and upon expanding in the soft parameter $\delta$,
\begin{equation}
k_{\alpha \ad}=\sum_{i=1}^{n-1}\frac{\partial^2}{\partial |i\r \partial [i|}+\frac{1}{\delta}\frac{\partial^2}{\partial |n\r \partial[n|} \, .
\end{equation}
Now note that $k$ annihilates arbitrary forms of tree-level stripped superamplitudes, in particular order by order in $\delta$. Applying $k$ to (\ref{tree-soft-theorem-bnd}) yields the constraint equations,
\begin{align}
 & \frac{\partial^2}{\partial |n\r\partial [n|}\left(S^{(0)}A^\tr_{n-1} \right)=0\label{conformal-constraint-lead} \, , \\
&\sum_{i=1}^{n-1}\frac{\partial^2}{\partial |i\r \partial [i|}\left(S^{(0)}A_{n-1}^\tr\right) +  \frac{\partial^2}{\partial |n\r \partial[n|}\left(S^{(1)}A_{n-1}^\tr\right)=0\label{conformal-constraint-sub} \, .
\end{align}
These equations allow us to determine the forms of the soft factors. In fact we shall require extra input from considerations of little group scaling, mass dimension and colour ordering. 
Firstly,  the soft operators must have mass dimension ${-}1$. 
Furthermore,  since we are taking a positive helicity particle soft, the soft operators must transform with weight ${-}2$ under the little group scaling,
\begin{equation}
|n\r \to t|n\r\ , \qquad |n] \to t^{-1}|n] \, ,
\end{equation}
and remain invariant under little group scaling for all other particles. Finally, since the amplitudes are colour ordered, the soft operators may only depend on particles $n-1$ and $1$ adjacent to $n$, since only these share a colour line with $n$. At one loop we will find that this simplifying assumption no longer holds since internal gluons carry colour dependence between arbitrary particles. Putting all this information together with the conformal Ward identities (\ref{conformal-constraint-lead}) and (\ref{conformal-constraint-sub}) suffices to determine $S^{(0)}$ and $S^{(1)}$ as written in (\ref{soft-op-leading}) and (\ref{soft-op-subleading}).\\\\
It is difficult to generalise the method of \cite{1405.2346} to  loop level, because 
the conformal anomaly takes a complicated form.
The current state of the art is restricted to MHV amplitudes and is rather intricate \cite{1002.1733}. By contrast, the dual conformal anomaly (\ref{dual-conformal-anomaly}) is very simple, which will allow us to make progress as we will see in the remainder of this section.

\subsection{Tree-Level Preliminaries}
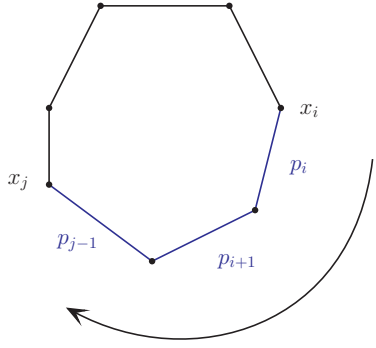
\begin{figure}[h]
\begin{center}
\scalebox{0.6}{
\fcolorbox{white}{white}{
  \begin{picture}(224,213) (170,-173)
    \SetWidth{1.0}
    \SetColor{Black}
    \Line(192,-27)(224,37)
    \Line(224,37)(304,37)
    \Line(304,37)(336,-27)
    \SetColor{Blue}
    \Line(336,-27)(320,-91)
    \Line(320,-91)(256,-123)
    \Line(256,-123)(192,-75)
    \SetColor{Black}
    \Line(192,-27)(192,-75)
    \Vertex(224,37){2}
    \Vertex(304,37){2}
    \Vertex(336,-27){2}
    \Vertex(320,-91){2}
    \Vertex(256,-123){2}
    \Vertex(192,-75){2}
    \Vertex(192,-27){2}
    \Text(349,-33)[lb]{\Large{\Black{$x_i$}}}
    \Text(167,-79)[lb]{\Large{\Black{$x_j$}}}
    \Text(343,-69)[lb]{\Large{\Blue{$p_i$}}}
    \Text(298,-128)[lb]{\Large{\Blue{$p_{i+1}$}}}
    \Text(198,-117)[lb]{\Large{\Blue{$p_{j-1}$}}}
    \Bezier(393,-59)(381,-166)(280,-194)(208,-155)\Line[arrow,arrowpos=1,arrowlength=10,arrowwidth=4,arrowinset=0.5](208.541,-155.291)(208,-155)%JaxoID:FBez[arrow,arrowpos=1,arrowlength=10,arrowwidth=4,arrowinset=0.5]
  \end{picture}
}
}
\end{center}
\caption{\it Solving for $x_j(p)$ clockwise around the polygon from $x_i$.}
\label{simple-polygon}
\end{figure}

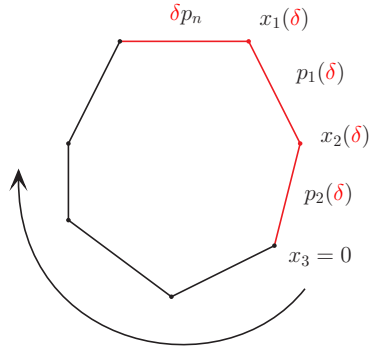
\begin{figure}[H]
\begin{center}
\scalebox{0.6}{
\fcolorbox{white}{white}{
  \begin{picture}(230,223) (155,-144)
    \SetWidth{1.0}
    \SetColor{Black}
    \Line(192,-17)(224,47)
    \SetColor{Red}
    \Line(224,47)(304,47)
    \Line(304,47)(336,-17)
    \Line(336,-17)(320,-81)
    \SetColor{Black}
    \Line(320,-81)(256,-113)
    \Line(256,-113)(192,-65)
    \Line(192,-17)(192,-65)
    \SetWidth{0.0}
    \Vertex(224,47){1.414}
    \SetColor{Red}
    \Vertex(304,47){1.414}
    \Vertex(336,-17){1.414}
    \SetColor{Black}
    \Vertex(320,-81){1.414}
    \Vertex(256,-113){1.414}
    \Vertex(192,-65){1.414}
    \Vertex(192,-17){1.414}
    \SetWidth{1.0}
    \Bezier(339,-108)(288,-173)(160,-146)(161,-39)\Line[arrow,arrowpos=1,arrowlength=10,arrowwidth=4,arrowinset=0.5](160.995,-39.801)(161,-39)%JaxoID:FBez[arrow,arrowpos=1,arrowlength=10,arrowwidth=4,arrowinset=0.5]
    \Text(256,58)[lb]{\Large{\Black{${\color{red}\delta} p_n$}}}
    \Text(335,20)[lb]{\Large{\Black{$p_1({\color{red}\delta})$}}}
    \Text(340,-56)[lb]{\Large{\Black{$p_2({\color{red}\delta})$}}}
    \Text(312,54)[lb]{\Large{\Black{$x_1({\color{red}\delta})$}}}
    \Text(350,-20)[lb]{\Large{\Black{$x_2({\color{red}\delta})$}}}
    \Text(330,-94)[lb]{\Large{\Black{$x_3 =0$}}}
  \end{picture}
  }
}
\end{center}
\caption{\it Setting $x_3=0$ is compatible with eliminating $|1]$ and $|2]$.}
\label{x30}
\end{figure}

Our goal is to constrain soft factors using dual  superconformal symmetry. We first work at tree level, and then extend the technique to one-loop amplitudes.
\\\\
To this end, we  begin by expanding the dual conformal boost generator (\ref{dual-conformal-boost-generator}) in powers of the soft parameter $\delta$. This involves solving for dual momenta $x_i$ in terms of momenta $p_j$. This procedure is ambiguous because of momentum conservation. 
In general we may freely fix any $x_i$ allowing us to perform the change of variables $x \to p$. More precisely, we determine $x_j$ as a sum of the $p_k$ between $x_i$ and $x_j$ as indicated in Figure \ref{simple-polygon}. The clockwise orientation is an arbitrary choice corresponding to taking $j > i$ cyclically.
\\\\ In Section \ref{supersoft-thms} we saw that momentum conservation is a subtle issue for subleading soft theorems. Therefore we must be careful regarding the ambiguity in base point $x_i$ and orientation around the polygon when solving for $x(p)$. In the following we use a prescription that eliminates a pair of antiholomorphic spinors $(|a],|b])$ according to the substitution (\ref{mom-cons-prescription}).\\\\
The simplest choice\footnote{In order to preserve momentum conservation and on-shell external momenta, a minimum of three momenta must acquire $\delta$ dependence, and hence a minimum of two dual momenta must be $\delta$-dependent. Our choice (\ref{x-defn})  achieves this.} is to fix $x_3=0$ and solve clockwise around the polygon, whence 
\begin{equation}
x_{k\alpha\dot\alpha}=-\sum_{j=3}^{k-1}|j]\l j|\, , \label{x-defn}
\end{equation}
for $k\neq 3$.  The solution (\ref{x-defn}) is compatible with eliminating $|1]$ and $|2]$. The only $\delta$-dependent region momenta are then $x_1$ and $x_2$ as shown in Figure \ref{x30}.
\noindent
Similarly,  for the fermionic variables we set $\theta_3 =0$ and write
\begin{equation}\label{theta-defn}
\l\theta_k^A|=-\sum_{j=3}^{k-1}\l j|\eta_j^A \, ,
\end{equation}
for $k\neq 3$. We should view (\ref{x-defn}) and (\ref{theta-defn}) as a frame choice well-adapted to the computations which follow. Of course, 
any result we derive in this frame may trivially be transformed to another using the substitution (\ref{mom-cons-prescription}).
\\\\
The soft expansion of the dual conformal boost generator is (with soft leg $n$)
\begin{equation}
\begin{aligned}
\label{dual-boost-exp}
K_{\a\ad}&=-\sum_{i\neq 3} \sum_{j=3}^{i-1}{}^{'} |j]\l i | \left(|j\r\cdot\frac{\partial}{\partial | i\r}\right) 
-\sum_{i\neq 2} \sum_{j=3}^i {}^{'} |i]\l j|\left(|j]\cdot\frac{\partial}{\partial |i]}\right)
-\sum_{i\neq 2} \sum_{j=3}^i {}^{'} |i]\l j|\eta_j^A\frac{\partial}{\partial \eta_i^A}
\nonumber
\\
&-\delta |n]\l 2|\left(|n\r \cdot\frac{\partial}{\partial |2\r}\right)-\delta|n]\l 1 |\left(|n\r\cdot\frac{\partial}{\partial | 1\r}\right)
-\delta|n]\l n|\left(|n]\cdot\frac{\partial}{\partial |n]}\right)
-\delta|1]\l n|\left(|n]\cdot\frac{\partial}{\partial |1]}\right) \nonumber
\\
& -\delta|n]\l n|\eta_n^A\frac{\partial}{\partial \eta_n^A}-\delta|1]\l n |\eta_n^A\frac{\partial}{\partial \eta_1^A} \, ,
\end{aligned}
\end{equation}
where $\sum_j\!\!{}^{'}$ indicates a sum over $j\neq n$. Similarly the statement  of dual conformal covariance (\ref{dual-conformal-covariance}) yields
\begin{equation}\label{covariance-expansion}
K_{\alpha\dot\alpha}A_n^\tr=\Big(\sum_{i\neq 3} \sum_{j=3}^{i-1} {}^{'} |j]\l j|+2\delta |n]\l n| \Big)A_n^\tr \, .
\end{equation}

\subsection{Leading Soft Theorems at Tree Level}\label{lead-soft-tree}
By keeping the leading $1/ \delta$ divergence in \eqref{dual-boost-exp} we find the following constraint equation for the leading soft factor, in analogy with (\ref{conformal-constraint-lead}), 
\begin{equation}\left( K_{\a \ad}A_n^\tr\right)_{\mathcal{O}(\delta^{-2})}=\left(K_{\a \ad}\right)_{\mathcal{O}(\delta^0)}\left(S^{(0)}A_{n-1}^\tr\right)=\Big(\sum_{i\neq 3} \sum_{j=3}^{i-1} {}^{'} |j]\l j|\Big)S^{(0)}A_{n-1}^\tr \, .
\label{basic-leading}\end{equation}
The covariance statement for $(n{-}1)$-point amplitudes gives 
\begin{equation}
\left(K_{\a \ad}\right)_{\mathcal{O}(\delta^0)}A_{n-1}^\tr=\Big(\sum_{i\neq 3,n} \sum_{j=3}^{i-1} {}^{'} |j]\l j|\Big)A_{n-1}^\tr \, .
\end{equation}
Hence (\ref{basic-leading}) simplifies to
\begin{equation}\left(K_{\a \ad}\right)_{\mathcal{O}(\delta^0)}S^{(0)}=\Big(\sum_{j=3}^{n-1}|j]\l j| \Big)S^{(0)}\, .
\label{master-leading}
\end{equation}
This leading order behaviour can be checked explicitly using the known form of $S^{(0)}$ and the formulae in Appendix \ref{useful-dual-conf-formulae}. 
Conversely we may use (\ref{master-leading}) to determine the form of $S^{(0)}$. Since our amplitudes are colour ordered, we may assume
\begin{equation}
S^{(0)}=f(\l a \ b\r,[a \ b]) \, ,
\end{equation}
where $a,b$ can take values in $\{n-1, n, 1\}$. To obtain the dual conformal transformation \eqref{master-leading} we must take
\begin{equation}
f\propto \frac{1}{\l n\ 1\r} \textrm{ or } \frac{1}{[n-1 \ n ]}\, .
\end{equation}
Note that the constant of proportionality must have mass dimension $0$. The constraint of little group scaling rules out the second option, and leads us uniquely to (\ref{soft-op-leading}).

\subsection{Subleading Soft Theorems at Tree Level}\label{sub-soft-tree}
At subleading order we employ the approach of  \cite{1405.1015}, allowing the freedom to use arbitrary forms of the stripped amplitudes in our derivations. The dual conformal analogue of (\ref{conformal-constraint-sub}) is
\begin{equation}
\begin{aligned}\left(K_{\a\ad}A_n^\tr\right)_{\mathcal{O}(\delta^{-1})}&=\left(K_{\a \ad}\right)_{\mathcal{O}(\delta^1)}\left(S^{(0)}A_{n-1}^\tr\right)+\left(K_{\a \ad}\right)_{\mathcal{O}(\delta^0)}\left(S^{(1)}A_{n-1}^\tr\right) 
\\&=2 |n]\l n| S^{(0)}A_{n-1}^\tr+\Big(\sum_{i\neq 3} \sum_{j=3}^{i-1} {}^{'} |j]\l j|\Big)S^{(1)}A_{n-1}^\tr \, . \label{basic-subleading}
\end{aligned}
\end{equation}
It is convenient to rewrite the first line of (\ref{basic-subleading}) to obtain
\begin{equation}
\begin{aligned}
\left(K_{\a\ad}A_n^\tr\right)_{\mathcal{O}(\delta^{-1})}
&=-\frac{|n]\l 1 |}{\l n \ 1 \r}A_{n-1}^\tr+S^{(0)}\left(K_{\a \ad}\right)_{\mathcal{O}(\delta^1)}A_{n-1}^\tr 
\\
&+\left[\left(K_{\a \ad}\right)_{\mathcal{O}(\delta^0)},S^{(1)}\right]A_{n-1}^\tr
+S^{(1)}\left[\Big(\sum_{i\neq 3,n} \sum_{j=3}^{i-1} {}^{'} |j]\l j|\Big)A_{n-1}^\tr\right] \, .
\end{aligned}
\end{equation}
Using the covariance statement for $(n{-}1)$-point amplitudes we get
\begin{eqnarray}
&&-\frac{|n]\l 1 |}{\l n \ 1 \r}A_{n-1}^\tr+S^{(0)}\left(K_{\a \ad}\right)_{\mathcal{O}(\delta^1)}A_{n-1}^\tr+\left[\left(K_{\a \ad}\right)_{\mathcal{O}(\delta^0)},S^{(1)}\right]A_{n-1}^\tr +A_{n-1}^\tr S^{(1)}\Big(\sum_{i\neq 3,n}\sum_{j=3}^{i-1} {}^{'}|j]\l j|\Big)
\nonumber
\\
&&=2|n]\l n| S^{(0)}A_{n-1}^\tr+\Big(\sum_{j=3}^{n-1}|j]\l j|\Big)S^{(1)}A_{n-1}^\tr \, .
\label{master-subleading}
\end{eqnarray}
We begin by verifying this using the known form of the subleading soft operator $S^{(1)}$. First note that
\begin{equation}
\frac{|n]\l 1 |}{\l n \ 1 \r}+S^{(1)}\Big(\sum_{i\neq 3,n} \sum_{j=3}^{i-1} {}^{'} |j]\l j|\Big)=2|n]\l n| S^{(0)} \, ,
\end{equation}
using a Schouten identity, whence (\ref{master-subleading}) becomes
\begin{equation}
-2\frac{|n]\l 1|}{\l n \ 1 \r}A_{n-1}^\tr+S^{(0)}\left(K_{\a \ad}\right)_{\mathcal{O}(\delta^1)}A_{n-1}^\tr+\left[\left(K_{\a \ad}\right)_{\mathcal{O}(\delta^0)},S^{(1)}\right]A_{n-1}^\tr = \Big(\sum_{j=3}^{n-1}|j]\l j|\Big)S^{(1)}A_{n-1}^\tr \, .
\label{simpler-subleading}
\end{equation}
Using formulae from Appendix \ref{useful-dual-conf-formulae} and Schouten identities we evaluate
\begin{equation}
\begin{aligned}
\left[\left(K_{\a \ad}\right)_{\mathcal{O}(\delta^0)},S^{(1)}\right]&=\Big(\sum_{j=3}^{n-1}|j]\l j|\Big)S^{(1)}+S^{(0)}|1]\l n|\left(|n]\cdot \frac{\partial}{\partial |1]}\right)+S^{(0)}|n]\l 1|\left(|n\r\cdot\frac{\partial}{\partial |1\r}\right)
\\&-\frac{|n]\l 1|}{\l n\ 1 \r}\left(|1\r\cdot\frac{\partial}{\partial |1\r}-|1]\cdot\frac{\partial}{\partial |1]}-\frac{\eta_1^A}{\l n \ 1 \r}\frac{\partial}{\partial \eta_1^A}\right) \, ,
\end{aligned}
\end{equation}
and observe that
\begin{align}\label{S0K0}
S^{(0)}\left(K_{\a \ad}\right)_{\mathcal{O}(\delta^1)}=-S^{(0)}|1]\l n| \left(|n]\cdot\frac{\partial}{\partial |1]}\right)-S^{(0)}|n]\l 1|\left(|n\r\cdot\frac{\partial}{\partial |1\r}\right) \, .
\end{align}
Hence (\ref{simpler-subleading}) simplifies to
\begin{equation}
\label{zzz}
-\frac{|n]\l 1|}{\l n\ 1 \r}\left(|1\r\cdot\frac{\partial}{\partial |1\r}-|1]\cdot\frac{\partial}{\partial |1]}-\frac{\eta_1^A}{\l n \ 1 \r}\frac{\partial}{\partial \eta_1^A}\right)A_{n-1}^\tr=2\frac{|n]\l 1|}{\l n \ 1 \r}A_n^\tr \, .
\end{equation}
On the  the left-hand side of  \eqref{zzz} we immediately recognise the appearance 
of the helicity operator (\ref{helicity-op}) for particle $1$. Recalling that superamplitudes have unit helicity completes the verification. \\\\
Conversely, we can use (\ref{master-subleading}) to derive the form of $S^{(1)}$. From Taylor series considerations it is natural to expect $S^{(1)}$ to be a derivative operator. We first split the constraint according to whether derivatives act, yielding
\begin{align}
S^{(0)}\left(K_{\a \ad}\right)_{\mathcal{O}(\delta^1)}+\left[\big(K_{\a \ad}\big)_{\mathcal{O}(\delta^0)},S^{(1)}\right] & =  \Big(\sum_{j=3}^{n-1}|j]\l j|\Big)S^{(1)} \, ,
\label{subl-tree-constraint-1}
\\
-\frac{|n]\l 1 |}{\l n \ 1 \r}+S^{(1)}\Big(\sum_{i\neq 3,n} \sum_{j=3}^{i-1} {}^{'} |j]\l j|\Big) & =  2|n]\l n| S^{(0)} \, .
\end{align}
Note that we might expect some mixing between the terms in each equation by virtue of the identity operator. The canonical representation of the identity under these circumstances is as a helicity operator. Hence we look for a form of $S^{(1)}$ which satisfies (\ref{subl-tree-constraint-1}) up to additive helicity operators.\\\\
The key observation is found by studying the derivative structure of (\ref{S0K0}). After Schoutening, the second term yields derivative structures which appear in $\left(K_{\a \ad}\right)_{\mathcal{O}(\delta^0)}$. Applying Occam's razor we postulate that such derivatives do not appear in $S^{(1)}$. On the contrary, the first term does not admit such a Schoutening since the derivative part involves antiholomorphic spinors. This leads us to the ansatz
\begin{equation}
S^{(1)}\sim |n]\cdot\frac{\partial}{\partial |1]}\, .
\end{equation}
The requirements of mass dimension and little group scaling suggest a prefactor $\l n\ 1 \r^{-1}$. Freedom to relabel the polygon in the opposite direction dictates the appearance of a similar term involving particle $(n-1)$. A similar process determines the fermionic terms, and we arrive at exactly (\ref{soft-op-subleading}). 
\\\\
Similarly to \cite{1405.2346} we can   fix the leading and subleading soft operators at tree level, with mild assumptions. More importantly, we may now use the simple form of the one-loop anomaly of dual conformal symmetry to study soft factorisation at one-loop level, as we will see below. Note in this context that the conventional conformal anomaly is much more complicated  and its general form is not known.

\subsection{One-Loop Preliminaries}
In Section \ref{supersoft-thms} we conjectured a form for the one-loop soft theorem
\begin{equation}
A_n^\lp \to  \frac{1}{\delta^2}\left(S^{(0)} A_{n-1}^\lp + S^{(0)\lp} A_{n-1}^\tr\right)+\frac{1}{\delta}\left(S^{(1)\tr}A_{n-1}^\lp + S^{(1)\lp}A_{n-1}^\tr\right) \, ,
\end{equation}
with previous results for $S^{(0)\lp}$ and the infrared-divergent part of $S^{(1)\lp}$ quoted in (\ref{S0lp}) and (\ref{S1lp}). We now derive dual conformal constraint equations on both $S^{(0)\lp}$ and $S^{(1)\lp}$ through $\mathcal{O} (\epsilon^0)$. These equations provide non-trivial checks on the known expressions. Furthermore, the one-loop subleading soft constraint suggests an ansatz for the hitherto unknown infrared-finite part of $S^{(1)\lp}$. 
\\\\
Paraphrasing (\ref{dual-conformal-anomaly}), the dual conformal operator acts on one-loop amplitudes to give 
\begin{equation}
K_{\alpha \ad} A_n^\lp
=\left(\textrm{anomaly}\right)A_n^\tr+\left(\textrm{covariance}\right)A_n^\lp \, .
\end{equation}
For later convenience we reproduce the soft expansion of the covariance statement from (\ref{covariance-expansion}),
\begin{equation}
\left(\textrm{covariance}\right)=\Big(\sum_{i\neq 3} \sum_{j=3}^{i-1} {}^{'} |j]\l j|+2\delta |n]\l n| \Big) \, .
\end{equation}
In the frame choice (\ref{x-defn}) the soft expansion of the anomaly (\ref{dual-conformal-anomaly}) is
\begin{equation}
\begin{aligned}
\left(\textrm{anomaly}\right)&=-\frac{2}{\epsilon}c_\Gamma \Big[\sum_{i\neq 1,3,n} \sum_{j=3}^{i-1} {}^{'} |j]\l j|\left(-\vphantom{p_i^j}(i-1 \ i)\right)^{-\epsilon}+ \sum_{j=3}^{n-1}|j]\l j|\left(-\vphantom{p_i^j}\delta(n-1 \ n)\right)^{-\epsilon} 
\\
&+\sum_{j=3}^{n-1}|j]\l j|\left(-\vphantom{p_i^j}\delta(n \ 1)  \right)^{-\epsilon}
+\delta |n]\l n|\left(-\vphantom{p_i^j}\delta(n \ 1)\right)^{-\epsilon}+\delta |n]\l n|\left(-\vphantom{p_i^j}(1 \ 2 )\right)^{-\epsilon}\Big] \, .
\end{aligned}
\end{equation}

\subsection{Leading Soft Theorems at One Loop}\label{sec:lead-soft-1-lp}
The one-loop version of (\ref{basic-leading}) is
\begin{equation}
\begin{aligned}
\left( K_{\a \ad}A_n^\lp\right)_{\mathcal{O}(\delta^{-2})}&=
\left(K_{\a \ad}\right)_{\mathcal{O}(\delta^0)}\left(S^{(0)}F^{(0)}A_{n-1}^\tr+S^{(0)}A_{n-1}^\lp\right)\notag
\nonumber \\
&=\left(\textrm{anomaly}\right)_{\mathcal{O}(\delta^0)}S^{(0)}A_{n-1}^\tr+\left(\textrm{covariance}\right)_\dzero S^{(0)}F^{(0)} A_{n-1}^\tr\notag\nonumber \\
&+\left(\textrm{covariance}\right)_\dzero S^{(0)}A_{n-1}^\lp \, ,
\label{basic-leading-lp}
\end{aligned}
\end{equation}
where $F^{(0)}$ is defined in (\ref{S0lp}).
This can be simplified significantly by recycling our tree-level knowledge; in fact, we can remove all terms involving the one-loop amplitude. Recall from (\ref{basic-leading}) that 
\begin{equation}
\left(K_{\a \ad}\right)_{\mathcal{O}(\delta^0)}\left(S^{(0)}A_{n-1}^\tr\right)=\left(\textrm{covariance}\right)_\dzero S^{(0)}A_{n-1}^\tr \, ,
\end{equation}
and hence we find that
\begin{equation}
\left(K_{\a \ad}\right)_{\mathcal{O}(\delta^0)} \left(S^{(0)} A_{n-1}^\lp\right)=\left(\textrm{covariance}\right)_\dzero S^{(0)}A_{n-1}^\lp+\left(\textrm{anomaly}\right)_{n-1}S^{(0)}A_{n-1}^\tr \, .
\end{equation}
Using these results (\ref{basic-leading-lp}) simplifies to
\begin{equation}\label{simple-leading-lp}
\left(K_{\a \ad}\right)_{\mathcal{O}(\delta^0)} F^{(0)}+\left(\textrm{anomaly} \right)_{n-1}=\left(\textrm{anomaly}\right)_{ \mathcal{O}(\delta^0)} \, ,
\end{equation}
or more explicitly, 
\begin{equation}\label{master-leading-lp}
\left(K_{\a \ad}\right)_{\mathcal{O}(\delta^0)}F^{(0)}=\frac{2}{\epsilon}c_\Gamma\Big(\sum_{j=3}^{n-1} |j]\l j|\Big)
\left[\vphantom{\frac{A}{B}}\left(-\vphantom{p_i^j}\delta(n-1 \ n)\right)^{-\epsilon}+\left(-\vphantom{p_i^j}\delta (n \ 1)\right)^{-\epsilon}-\left(-\vphantom{p_i^j}(n-1 \ 1 )\right)^{-\epsilon}\right] \, .
\end{equation}
Firstly we wish to verify  that (\ref{master-leading-lp}) holds using the known expression for $F^{(0)}$ in (\ref{S0lp}). It is easy to see that this is true at $\mathcal{O}(\epsilon^{-1})$. 
Using results from Appendix \ref{useful-dual-conf-formulae} 
we find that
\begin{equation}\label{leading-lp-calc}
\left(K_{\a \ad}\right)_{\mathcal{O}(\delta^0)}\left(\frac{(n-1 \ 1)}{(n-1 \ n)(n \ 1)}\right)= 2\Big(\sum_{j=3}^{n-1}|j]\l j|\Big)\left(\frac{(n-1 \ 1 )}{(n-1 \ n )  (n \ 1 )}\right) \, ,
\end{equation}
whence at $\mathcal{O}(\epsilon^0)$ in (\ref{master-leading-lp}) both sides evaluate to
\begin{equation}
2c_\Gamma\Big(\sum_{j=3}^{n-1}|j]\l j|\Big)\log\left(-\frac{1}{\delta^2}\frac{(n-1 \ 1)}{(n-1 \ n)(n \ 1 )}\right) \, ,
\end{equation}
confirming the consistency of (\ref{master-leading-lp}) at $\mathcal{O}(\epsilon^0)$ also.\\\\
Conversely we can use (\ref{master-leading-lp}) as a constraint equation to determine $F^{(0)}$ up to and including $\epsilon^0$ terms, provided that we assume that $F^{(0)}$ only depends on particles $n-1$, $n$ and $1$ and is a dimensionless, helicity-blind function. The derivation proceeds analogously to that in Section \ref{lead-soft-tree}.
\\\\ 
Na\"\i vely, the restriction to particles neighbouring $n$ seems unreasonable from the Wilson loop perspective. Indeed, we might expect contributions from diagrams where an internal gluon connects an arbitrary edge to $p_n$. However, the scalar boxes corresponding to the non-cusp diagrams do not contribute in the leading soft limit. This is perhaps most obvious from the perspective of MHV diagrams \cite{hep-th/0510253}.

\subsection{Subleading Soft Behaviour at One Loop}\label{sec:sub-soft-1-lp}
The one-loop version of (\ref{basic-subleading}) is
\begin{equation}
\begin{aligned}
\left(\K_{\a\ad}A_n^\lp\right)_{\mathcal{O}(\delta^{-1})}&=\left(\K_{\a \ad}\right)_{\mathcal{O}(\delta^1)}\left(S^{(0)}F^{(0)}A_{n-1}^\tr+S^{(0)}A_{n-1}^\lp\right)\\ &+\left(K_{\a \ad}\right)_{\mathcal{O}(\delta^0)}\left(S^{(1)}A_{n-1}^\lp + F^{(1)}S^{(1)}A_{n-1}^\tr+ZA_{n-1}^\tr\right)
\\&=\left(\textrm{anomaly}\right)_{\mathcal{O}(\delta^1)}S^{(0)}A_{n-1}^\tr+\left(\textrm{covariance}\right)_\done S^{(0)}F^{(0)} A_{n-1}^\tr\\
&+\left(\textrm{covariance}\right)_\done S^{(0)}A_{n-1}^\lp+\left(\textrm{anomaly}\right)_\dzero S^{(1)}A_{n-1}^\tr\\
&+\left(\textrm{covariance}\right)_\dzero S^{(1)} A_{n-1}^\lp + \left(\textrm{covariance}\right)_\dzero F^{(1)}S^{(1)} A_{n-1}^\tr\\
&+\left(\textrm{covariance}\right)_\dzero Z A_{n-1}^\tr \, ,
\label{master-subleading-lp}
\end{aligned}
\end{equation}
where $F^{(1)}$ and $Z$ are defined in (\ref{subsoftZ}).
Just as in Section \ref{sec:lead-soft-1-lp} we can remove all terms involving $A_{n-1}^\lp$ by recycling tree-level knowledge. Recall from (\ref{basic-subleading}) that
\begin{equation}
\begin{aligned}
&\left(\K_{\a\ad}\right)_\done\left(S^{(0)}A_{n-1}^\tr\right)+\left(\K_{\a\ad} \right)_\dzero\left(S^{(1)}A_{n-1}^\tr\right)
 \\
 &=
\left(\textrm{covariance}\right)_\done S^{(0)}A_{n-1}^\tr +\left(\textrm{covariance}\right)_\dzero \ S^{(1)}A_{n-1}^\tr \, ,
\end{aligned}
\end{equation}
and hence that
\begin{equation}
\begin{aligned}
&
\left(\K_{\a\ad}\right)_\done\left(S^{(0)}A_{n-1}^\lp\right)+\left(\K_{\a\ad} \right)_\dzero\left(\  S^{(1)}A_{n-1}^\lp\right)
 \\
&=\left(\textrm{covariance}\right)_\done S^{(0)}A_{n-1}^\lp %\\*
+ \left(\textrm{covariance}\right)_\dzero  \  S^{(1)}A_{n-1}^\lp
+ S^{(1)}\left[\left(\textrm{anomaly}\right)_{n-1}A_{n-1}^\tr\right] \, .
\end{aligned}
\end{equation}
Applying these results to (\ref{master-subleading-lp}) we get
\begin{equation}
\begin{aligned}
\label{simple-subleading-lp}
&
S^{(0)}A_{n-1}^\tr\left(\K_{\a \ad}\right)_{\mathcal{O}(\delta^1)}F^{(0)}+(F^{(0)}-F^{(1)})\left(\K_{\a \ad}\right)_{\mathcal{O}(\delta^1)}\left(S^{(0)}A_{n-1}^\tr\right)
 \\
&
+\  S^{(1)}A_{n-1}^\tr\left(\K_{\a \ad}\right)_{\mathcal{O}(\delta^0)}F^{(1)}
+A_{n-1}^\tr\left(\K_{\a \ad}\right)_{\mathcal{O}(\delta^0)}Z 
 \\
&=
(F^{(0)}-F^{(1)})\left(\textrm{covariance}\right)_\done S^{(0)}A_{n-1}^\tr+\left(\textrm{anomaly}\right)_{\mathcal{O}(\delta^1)}S^{(0)}A_{n-1}^\tr
 \\
&
+\left(\textrm{anomaly}\right)_\dzero \  S^{(1)}A_{n-1}^\tr
+\left(\textrm{anomaly}\right)_{n-1} S^{(1)}A_{n-1}^\tr-
 S^{(1)}\left[\left(\textrm{anomaly}\right)_{n-1}\right]A_{n-1}^\tr
  \\
 &
 +\left[\left(\textrm{covariance}\right)_\dzero-\left(\textrm{covariance}\right)_{n-1} \right]ZA_{n-1}^\tr \, .
\end{aligned}
\end{equation}
To proceed, we separate this result into two equations, depending on whether derivatives act on $A_{n-1}^\tr$; of course there may be some cancellations between these equations via the appearance of helicity operators. With this separation we have derivative terms,
\begin{equation}
\begin{aligned}
\label{deriv-terms}
&
(F^{(0)}-F^{(1)})S^{(0)}\left(\K_{\a \ad}\right)_{\mathcal{O}(\delta^1)}A_{n-1}^\tr+\ S^{(1)}A_{n-1}^\tr\left(\K_{\a \ad}\right)_{\mathcal{O}(\delta^0)}F^{(1)}
 \\
&=\left(\textrm{anomaly}\right)_\dzero \  S^{(1)}A_{n-1}^\tr
-\left(\textrm{anomaly}\right)_{n-1} S^{(1)}A_{n-1}^\tr \, ,
\end{aligned}
\end{equation}
and non-derivative terms,
\begin{equation}
\begin{aligned}
\label{non-deriv-terms}
&
S^{(0)}A_{n-1}^\tr\left(\K_{\a \ad}\right)_{\mathcal{O}(\delta^1)}F^{(0)}+(F^{(0)}-F^{(1)})A_{n-1}^\tr\left(\K_{\a \ad}\right)_{\mathcal{O}(\delta^1)}S^{(0)}
+A_{n-1}^\tr\left(\K_{\a \ad}\right)_{\mathcal{O}(\delta^0)}Z 
 \\
&
=(F^{(0)}-F^{(1)})\left(\textrm{covariance}\right)_\done S^{(0)}A_{n-1}^\tr+\left(\textrm{anomaly}\right)_{\mathcal{O}(\delta^1)}S^{(0)}A_{n-1}^\tr
 \\
&
-S^{(1)}\left[\left(\textrm{anomaly}\right)_{n-1}\right]A_{n-1}^\tr
 +\left[\left(\textrm{covariance}\right)_\dzero-\left(\textrm{covariance}\right)_{n-1} \right]ZA_{n-1}^\tr \, .
\end{aligned}
\end{equation}
We focus first on equation (\ref{deriv-terms}). Note that the derivatives in $\left(\K_{\a \ad}\right)_{\mathcal{O}(\delta^1)}$ and $S^{(1)}$ do not combine to yield a helicity operator. Therefore we may assume that this equation is truly decoupled from (\ref{non-deriv-terms}). Using (\ref{simple-leading-lp}) we see that (\ref{deriv-terms}) is satisfied if we choose $F^{(1)}=F^{(0)}$. This is consistent with the known infrared divergent behaviour of $F^{(1)}$ and extends it to finite order in $\epsilon$. 
\\\\
With this choice, (\ref{non-deriv-terms}) simplifies to give
\begin{equation}
\begin{aligned}
\label{simplest-subl-lp}
&
S^{(0)}A_{n-1}^\tr\left(\K_{\a \ad}\right)_{\mathcal{O}(\delta^1)}F^{(0)}
+A_{n-1}^\tr\left(\K_{\a \ad}\right)_{\mathcal{O}(\delta^0)}Z =\left(\textrm{anomaly}\right)_{\mathcal{O}(\delta^1)}S^{(0)}A_{n-1}^\tr
 \\
&
-S^{(1)}\left[\left(\textrm{anomaly}\right)_{n-1}\right]A_{n-1}^\tr
 +\left[\left(\textrm{covariance}\right)_\dzero-\left(\textrm{covariance}\right)_{n-1} \right]ZA_{n-1}^\tr \, .
\end{aligned}
\end{equation}
Thus we have arrived at the dual conformal constraint equation on the one-loop subleading soft anomaly promised in Section \ref{main-summary}.
\subsubsection*{Constraint on the Infrared-Divergent Anomaly}
We now expand in $\epsilon$ to find constraint equations for $Z$ at each order. We write
\begin{equation}
Z=\frac{1}{\epsilon^2}Z_{-2}+\frac{1}{\epsilon}Z_{-1}+Z_{0}+\mathcal{O}(\epsilon) \, .
\end{equation}
At leading order in $\epsilon$ the anomaly constraint (\ref{simplest-subl-lp}) becomes
\begin{equation}\label{leading-eps}
\left(\K_{\a \ad}\right)_{\mathcal{O}(\delta^0)}Z_{-2} =\left[\left(\textrm{covariance}\right)_\dzero-\left(\textrm{covariance}\right)_{n-1} \right]Z_{-2}=\Big(\sum_{j=3}^{n-1}|j]\l j|\Big) Z_{-2} \, .
\end{equation}
Clearly this is consistent with the choice $Z_{-2}=0$ implicit in (\ref{S1lp}). For the converse argument, first note that (\ref{leading-eps}) has exactly the same form as (\ref{master-leading}). We therefore employ logic similar to the leading tree-level case. Indeed, since we are dealing with infrared-divergent terms, 
\begin{equation}
Z_{-2}=f(\l a \ b \r,[a \ b ]) \, ,
\end{equation}
where $a,b$ takes values in $\{n-2,n-1,n,1,2\}$ by the Wilson loop observations of Section \ref{supersoft-thms}. Following Section \ref{lead-soft-tree}, we see that if $Z_{-2}\neq 0$ then we must have $Z_{-2} = S^{(0)}$. But reinserting factors of $\delta$ shows that $S^{(0)}$ can only appear as a leading soft divergence. Hence the constraint equation fixes 
\begin{equation}
Z_{-2} =0 \, .
\end{equation}
At subleading order in $\epsilon$ the anomaly constraint (\ref{simplest-subl-lp}) becomes
\begin{equation}
\begin{aligned}
\label{order-eps-1-constraint}
&
\Big[\left(\K_{\a \ad}\right)_{\mathcal{O}(\delta^0)}-\sum_{j=3}^{n-1}|j]\l j|\Big]Z_{-1}
 \\
&
=-|1]\l 1|\frac{[n-1 \ n ]}{[n-1 \ 1 ]\l n \ 1 \r}+\frac{|n]\l n-1|}{\l n-1 \ n \r}-\frac{|n]\l 1|}{\l n \ 1 \r}+|n-1]\l n-1|\frac{[n \ 1 ]}{\l n-1 \ n \r[n-1\ 1]} \, .
\end{aligned}
\end{equation}
Note that this is symmetric under relabelling the polygon anticlockwise, as expected. From (\ref{S1lp}) we have \cite{1405.1015}
\begin{equation}\label{Sllp2}
Z_{-1}=\frac{[n-1 \ n]}{[n-1 \ 1 ]\l 1 \ n \r}+\frac{[2 \ n ]}{[2 \ 1 ]\l 1 \ n \r}-\frac{[1 \ n]}{[1 \ n-1]\l n-1 \ n \r}-\frac{[n-2 \ n]}{[n-2 \ n-1]\l n-1 \ n \r} \, .
\end{equation}
The reader may verify that this satisfies the constraint equation, using formulae from Appendix \ref{useful-dual-conf-formulae}. Conversely we write an ansatz,
\begin{equation}
Z_{-1}=g(\l a \ b \r,[a \ b ]) \, ,
\end{equation}
where $a,b$ can take values in $\{n-2,n-1,n,1,2\}$ by the same logic as for $Z_{-2}$. We assume for simplicity that each term on the right-hand side of (\ref{order-eps-1-constraint}) emerges from a single term in $Z_{-1}$. Then (\ref{usefulKspartIfirst})--(\ref{usefulKspartIlast}) immediately suggest the result (\ref{Sllp2}), which is clearly consistent with spinor weight and dimension constraints.
\subsubsection*{Constraint on the Infrared-Finite Anomaly}
Finally we consider  the  $\mathcal{O}(\epsilon^0)$ terms. The anomaly constraint (\ref{simplest-subl-lp}) becomes
\begin{equation}
\begin{aligned}
\label{Z0-constraint}
&\Big[\left(\K_{\a \ad}\right)_{\mathcal{O}(\delta^0)}-\sum_{j=3}^{n-1}|j]\l j|\Big]Z_0
 \\ 
&
=2\Big[ Z_{-1} \sum_{j=3}^{n-1}|j]\l j|  -|n-1]\l n-1| \frac{[n-2 \ n]}{\l n-1 \ n\r [n-1 \ n-2]}+|1]\l 1|\frac{[2 \ n]}{[2 \ 1 ]\l 1 \ n \r}\Big]\hphantom{fillerfiller}
 \\
&
+\left[\frac{|n]\l 1|}{\l n \ 1 \r}+2\frac{|n]\l n-1|}{\l n-1 \ n \r}-|1]\l n|\frac{\l n-1 \ 1 \r[n\ n-1]}{\l n-1 \ n \r \l n \ 1 \r [1 \ n-1]}\right]
\log\left(-\frac{(n-1 \ 1 )}{(n-1 \ n)(n \ 1 )}\right)
 \\
&
-2\frac{|n]\l n-1|}{\l n-1 \ n \r}\log(-(n-1 \ 1 ))+2|n]\l n|\frac{\l n-1 \ 1 \r}{\l n-1 \ n \r\l n \ 1 \r}\log(-(n \ 1) ) \, .
\end{aligned}
\end{equation}
We now employ this formula to find plausible coefficients for the $\log \delta$ terms appearing in $Z_0$. The constraint equation (\ref{Z0-constraint}) immediately suggests that these take the form,%
\footnote{Recall that in the soft limit $p_n \to \delta p_n$.}
\begin{equation}
A\log(-(n-1\ n))+B\log(-(n \ 1)) \, .
\end{equation}
Relabelling symmetry ensures that it suffices to predict coefficient $A$. We make the ansatz,
\begin{equation}
A = S^{(0)}h(\l a \ b \r,[a \ b ]) \, ,
\end{equation}
where $a,b$ take values in $\{n-2,n-1,n,1,2\}$. Unlike the $Z_{-2}$ and $Z_{-1}$ cases considered above, there is no rigorous argument for this assumption, since the $\log\delta$ terms do not only emerge from cusp diagrams. Nevertheless it seems plausible to expect that such divergent terms only involve particles close to $n$. In Section \ref{verification} we shall see that this ansatz holds for MHV amplitudes, but not in the NMHV sector.
\\\\
A general one-loop amplitude in $\mathcal{N}=4$ SYM theory involves functions of transcendentality $2$. Therefore we expect the soft anomaly $Z_0$ to contain functions of transcendentality $1$ and $0$. We may hence deduce from (\ref{Z0-constraint}) a constraint on $h$ by examining the coefficient of $\log(-(n-1 \ n ))$, namely
\begin{equation}
\left(\K_{\a \ad}\right)_{\mathcal{O}(\delta^0)}h=-|n-1] \l n-1 | \frac{(n \ 1 )}{(n-1 \ 1)}- |n] \l n|  +|1] \l 1| \frac{(n-1 \ n)}{(n-1 \ 1)} \, .
\end{equation}
Equations (\ref{usefulKspartIIfirst}) and (\ref{usefulKspartIIlast}) hence suggest that
\begin{equation}\label{predictedlogsoft}
Z_0|_{\log\delta}= \left(\frac{( n \ 1 )}{(n-1 \ 1)}+\frac{(n-2 \ n)}{(n-2 \ n-1 )}-\frac{(n- 2 \ 1)(n-1 \ n )}{(n-2 \ n-1)(n-1 \ 1)}\right)S^{(0)}\log(-(n-1 \ n)) + (i \leftrightarrow n-i) \, .
\end{equation}
%a form clearly compatible with dimension and spinor weight requirements. 
We now proceed to verify this prediction by explicitly computing the subleading soft anomaly in the MHV and NMHV sectors. Beware that $Z_0$ itself does not suffice to reconstruct the subleading soft behaviour of an $n$-point amplitude; we must also remember feed-down terms (see the discussion near \eqref{feed-down}). We consider this nicety in detail  in Section \ref{mhv-twistors}. 
%%%%%%%%%%

\section{Extracting the Infrared-Finite Soft Behaviour  at One Loop}\label{verification}
\subsection{Summary of Results}\label{summary-verification}
In this section we determine the subleading soft contribution for $n$-point one-loop MHV amplitudes and for six-point and seven-point one-loop NMHV amplitudes. We first present the subleading soft behaviour of some low-point MHV cases, extracted via the unitarity method. We then use momentum twistor technology to derive a surprisingly compact expression for the subleading soft term at $n$-point modulo $A_n^\tr$, namely%
\footnote{See Appendix \ref{momentum-twistors} for a crash review of momentum twistors. Round brackets such as $(6)$ appearing below represent the dual superconformal $R$-invariants, and are defined in  \eqref{rrr} and \eqref{rb}. }

\begin{equation}
\begin{aligned}
\label{111}
&
\frac{\l n-1 \ 1 \r}{\l n-1 \ n \r}\sum_{j=4}^{n-4}\log\left(\frac{y_{n-1j}^2}{y_{1j}^2}\right)\frac{\l n-2 \ n-1 \ j-1 \ j \r\l n-2 \ n-1 \ n \ 1 \r}{\l n-2 \ n-1 \ 1 \ j-1 \r \l n-2 \ n-1 \ 1 \ j \r}
 \\
&
+\frac{\l n-1 \ 1 \r}{\l n \ 1 \r}\sum_{j=5}^{n-3}\log\left(\frac{y_{2j}^2}{y_{1j}^2}\right)\frac{\l n-1 \ n \ 1 \ 2 \r\l j-1 \ j \ 1 \ 2 \r}{\l n-1 \ 1 \ 2 \ j \r\l n-1 \ 1 \ 2 \ j-1 \r}+\textrm{boundary terms} \, ,
\end{aligned}
\end{equation}
The boundary terms have a universal form for all $n\geq 7$. In particular, the $\log \delta $ dependence is simply
\begin{equation}
\left(\frac{(n \ 1) + (n \ 2)}{(1 \ 2 )}-\frac{s_{n-1,1,2}(n \ 1 )}{(n-1 \ 1)(1 \ 2)}\right)\log(-(n \ 1))+( i \leftrightarrow n-i ) \, .
\end{equation}
We finally investigate the possibility of universality carrying over to NMHV amplitudes by identifying the subleading $\log\, \delta$ terms in low-point cases. Again intricate cancellations yield a remarkably simple result, but of a slightly different form to the MHV sector. Explicitly we find terms at six and seven points,
\begin{gather}
\frac{1}{2}\frac{\l n-1 \ 1 \r}{\l n-1 \ n \r}(6)\frac{ \l 2 \ 3 \ 4 \ 5 \r\l 4 \ 5 \ 6 \ 1 \r}{\l 1 \ 2 \ 4 \ 5 \r\l 3 \ 4 \ 5 \ 1 \r}\log(-(6 \ 1))+( i \leftrightarrow n-i )\, ,
\\
\frac{1}{2}\frac{\l n-1 \ 1 \r}{\l n-1 \ n \r}\left[(5)\frac{\l 2 \ 3 \ 4 \ 6 \r \l 5 \ 6 \ 7 \ 1 \r}{\l 3 \ 4 \ 6 \ 1 \r \l 1 \ 2 \ 5 \ 6 \r}+(3)\frac{\l 2 \ 4 \ 5 \ 6 \r\l 5 \ 6 \ 7 \ 1 \r}{\l 4 \ 5 \ 6 \ 1 \r\l 1 \ 2\ 5 \ 6 \r}\right]\log(-(7 \ 1 ))+ ( i \leftrightarrow n-i )\, ,
\label{222}
\end{gather}
respectively. We conjecture that the $\log \delta$ terms display universal behaviour for arbitrary $n$ within each $\textrm{N}^k\textrm{MHV}$ sector, but not between different sectors.
\\\\
Throughout this section we employ the approach 
of \cite{1405.1015}, with a symmetric momentum conservation prescription eliminating $|n-1]$ and $|1]$. In particular this implies that the feed-down terms from Taylor-expanding  $S^{(0)}A_{n-1}^\lp$  in the soft parameter exactly cancel the contribution from $S^{(1)}A_{n-1}^\lp$. Therefore the form of the lower-point amplitude becomes irrelevant to the calculation of the subleading soft anomaly. We also neglect a factor of $c_\Gamma$ for brevity -- this may be reinserted easily afterwards.

\subsection{MHV Amplitudes via Unitarity}
All planar one-loop MHV amplitudes in $\mathcal{N}=4$ SYM theory may be written as \cite{hep-ph/9403226}
\begin{equation}
\mathcal{A}_n^\lp=\mathcal{A}_n^\tr\sum_{\textrm{channels}} F^{\textrm{2me}}\, ,
\end{equation}
where $F^{\textrm{2me}}$ is a (possibly degenerate) two-mass easy box function. Generically, 
\begin{equation}
\begin{aligned}
F^{2\textrm{me}}(K,L)&=-\frac{1}{\ep^2}\left[(-s)^{-\ep}+(-t)^{-\ep}-(-K^2)^{-\ep}-(-L^2)^{-\ep}\right]+\mathrm{Li}_2\left(1-\frac{K^2}{s}\right)
 \\
&+\mathrm{Li}_2\left(1-\frac{K^2}{t}\right)+\mathrm{Li}_2\left(1-\frac{L^2}{s}\right)+\mathrm{Li}_2\left(1-\frac{L^2}{t}\right)-\mathrm{Li}_2\left(1- \frac{K^2 L^2}{s t}\right)+\frac{1}{2}\log^2\left(\frac{s}{t}\right)\, ,
\end{aligned}
\end{equation}
while, for  degenerate cases, 
\begin{gather}
\begin{align}
F^{0\textrm{m}}&=-\frac{1}{\ep^2}\left[(-s)^{-\ep}+(-t)^{-\ep}\right]+\frac{1}{2}\log^2\left(\frac{s}{t}\right)+\frac{\pi^2}{2} \, , \\
F^{1\textrm{m}}(K)&=-\frac{1}{\ep^2}\left[(-s)^{-\ep}+(-t)^{-\ep}-(-K^2)^{-\ep}\right]+\mathrm{Li}_2\left(1-\frac{K^2}{s}\right)+\mathrm{Li}_2\left(1-\frac{K^2}{t}\right)+\frac{1}{2}\log^2\left(\frac{s}{t}\right)+\frac{\pi^2}{6} \, ,
\end{align}
\end{gather}
where $K$ and $L$ denote the momenta of massive corners and $(s,t)$ are defined by $(\textrm{vertical},\textrm{horizontal})$ cuts respectively. 
\\\\
To calculate the subleading soft behaviour, we must in principle Taylor expand all box functions. Many may be immediately discarded, along the lines outlined in Section \ref{supersoft-thms}. Specifically the only nonzero terms emerge from boxes corresponding to Wilson loop diagrams in which the internal gluon ends on particle lines $n -1$, $n$ or $1$.

\subsubsection*{Five-Point Amplitude}
The simplest non-trivial subleading soft behaviour appears at five points. In this case, the two-mass easy boxes degenerate to one-mass boxes. The subleading soft five-point infrared-finite term divided by $A_5^\tr$ is 
\begin{equation}\label{5pt-subl}
-\frac{\l 4\ 5\r  [ 2\ 5]  }{\l 3\ 4\r  [ 2\ 3]
   }\log \left(\frac{(4 \ 5)}{(1 \ 5)(2 \ 3)  }\right)-\frac{\l 1\ 5\r  [ 3\ 5]  }{\l 1\ 2\r  [ 2\ 3] }\log \left(\frac{(1 \ 5)}{(2 \ 3)(4 \ 5)}\right)\, ,
\end{equation}
For compactness we have implicitly recombined terms using four-point momentum conservation where appropriate. We have checked this result numerically using the Mathematica package documented in Appendix \ref{package}.
\\\\
The simplifications required to reach (\ref{5pt-subl}) involve intricate cancellations between roughly $20$ terms from different boxes. This suggests that box functions are poorly adapted to the calculation of subleading soft behaviour.

\subsubsection*{Six-Point Amplitude}
At six points we discover new structure associated with the appearance of non-degenerate two-mass boxes. The subleading soft infrared-finite contribution modulo $A_6^\tr$ is
\begin{align*}
\label{6pt-subl}
&
\frac{(3 \ 4)(1 \ 6)}{(1 \ 2)  (1 \ 5)}\left[1+\log \left(\frac{(1 \ 2) (1 \ 5)  }{(1 \ 6)(3 \ 4) }\right)\right]+\frac{(2 \ 3)
   (5 \ 6) }{(4 \ 5)(1 \ 5 )  }\left[1+\log \left(\frac{(1 \ 5 )(4 \ 5) }{(2 \ 3)(5 \ 6) }\right)\right]
    \\
&
   -\frac{\l 1 \ 6\r[ 3\ 4](\l 3 \ 4\r[4 \ 6] + \l 3 \ 5\r[5 \ 6]) }{(1 \ 2)  \l 1\ 5\r  [ 4\ 5] }\log (-(1 \ 2))-\frac{\l 5 \ 6 \r[ 2 \ 3 ](\l 1 \ 3 \r[1 \ 6] + \l 2 \ 3 \r [2 \ 6])}{(4 \ 5)\l 1\ 5\r   [ 1\ 2]  }\log (-(4 \ 5) )
    \\
   &
+\frac{  ((1 \ 6)+(2 \ 6)) }{(1 \ 2)} \left[\log
   \left(-\frac{(1 \ 6)(3 \ 4)}{(1 \ 5) }\right)-1\right]
   +\frac{  ((4 \ 6) +(5 \ 6) ) }{(4 \ 5) }\left[\log \left(-\frac{(2 \ 3) (5 \ 6) }{(1 \ 5)
   }\right)-1\right] \\
   &
+\frac{[ 2\ 6]\l 5 \ 6\r  }{\l 1\ 5\r  [ 1\ 2] }\log \left(-\frac{(1 \ 5)(2 \ 3) }{(3 \ 4)}\right)+\frac{[ 4\ 6]\l 1 \ 6 \r  }{\l 1\ 5\r  [ 4\ 5] }\log \left(-\frac{(1 \ 5)(3 \ 4) }{(2 \ 3) }\right) \numberthis \, .
\end{align*}
As in the five-point case, very non-trivial simplifications take place --  the Taylor expansion initially produces some $200$ terms. Mathematica numerics exactly confirm our concise formula.
\\\\
It is instructive to perform a consistency check that the six-point result (\ref{6pt-subl}) reproduces the five-point result (\ref{5pt-subl}) when we make particle $3$ soft. Taking the limit and relabelling appropriately we obtain
\begin{multline}\label{5pt-from-6pt}
\left(\frac{\l 4 \ 5 \r[ 2\ 5] }{\l 1\ 4\r  [ 1\ 2] }+\frac{\l 1 \ 5 \r[ 3\ 5]}{\l 1\ 4\r  [ 3\ 4] }\right)\log (-(1 \ 4))+\frac{(1 \ 5)  +(2 \ 5)}{(1 \ 2) }\log (-(1 \ 5) )+\frac{(3 \ 5)+(4 \ 5) }{(3 \ 4) }\log (-(4 \ 5) )\, .
\end{multline}
Na\"\i vely it looks impossible to equate (\ref{5pt-from-6pt}) and (\ref{5pt-subl}), however we only require them to match when a consistent momentum conservation prescription is applied to both. The relatively simple form of (\ref{5pt-subl}) is a consequence of the special four-point kinematics,
$
[1 \ 2] = \l 3 \ 4 \r [2 \ 3 ]    /  \l 1 \ 4\r \, .
$
\subsubsection*{$\mathrm{log} \, \delta $ Terms}
To complete our analysis we concentrate  on terms involving a  $\log \delta $. Recent evidence \cite{1503.05156} shows that these may be universal in QCD processes. Indeed these terms are truly infrared divergent, so intuitively one might expect enhanced universality to ensure such quantities cancel in any physical observable. At five points we have from (\ref{5pt-from-6pt})
\begin{equation}\label{logdel5}
\frac{(1 \ 5)+(2 \ 5)}{(1 \ 2)} \log (-(1 \ 5) )
+\frac{(3 \ 5)+(4 \ 5)  }{(3 \ 4) }\log (-(4 \ 5) )\, ,
\end{equation}
while at six points (\ref{6pt-subl}) yields
\beq
\label{logdel6}
%&&
\frac{(1 \ 6)+(2 \ 6)}{(1 \ 2)} \log (-(1 \ 6 ))
+\frac{(4 \ 6)+(5 \ 6)  }{(4 \ 5) }\log (-(5 \ 6) )
%\nonumber \\
%&&
-\frac{(3 \ 4 )(1 \ 6)}{(1 \ 2)(1 \ 5)}\log(-(1 \ 6))-\frac{(2 \ 3)(5 \ 6)}{(1 \ 5)(4 \ 5)}\log(-(5 \ 6))\, .
\eeq
The chances of a simple universal result look slim based on this evidence. In (\ref{logdel6}) new structures appear, in addition to a generalisation of (\ref{logdel5}). However, we shall see in Section \ref{mhv-twistors} that the complexity of $\log\delta$ terms does not grow with particle number in general. From the perspective of box functions, this is reasonable: a new type of box function enters at six points, after which no further new functions appear in the MHV sector.

\subsection{MHV Amplitudes via Momentum Twistors}\label{mhv-twistors}
We saw in Section \ref{1-lp-soft} that two classes of Wilson loop diagrams contribute to the subleading soft behaviour. Cusp diagrams give rise to the infrared-divergent piece of any one-loop SYM amplitude,
\begin{equation}\label{ir-div-pieces}
-\frac{1}{\epsilon^2} \sum_{i=1}^n \left(-\vphantom{p^i_j}(i \ i+1)\right)^{-\epsilon}\ .
\end{equation}
Non-cusp diagrams with an internal gluon ending on at least one $\delta$-dependent edge also feature. In the MHV sector these correspond to the finite parts of the two-mass easy boxes in Figure \ref{generic-2me}. Note that $i$ and $j$ must be separated by at least one intervening particle cyclically. For the symmetric momentum conservation prescription eliminating $(|n-1],|1])$, we can restrict to diagrams where $i$ or $j$ is in $\{n-1,n,1\}$.
\begin{figure}[H]
\begin{center}
\scalebox{0.8}{
\fcolorbox{white}{white}{
  \begin{picture}(208,193) (238,-48)
    \SetWidth{1.0}
    \SetColor{Black}
    \Line(288,81)(288,1)
    \Line(288,81)(368,81)
    \Line(368,81)(368,1)
    \Line(288,1)(368,1)
    \Line(240,97)(288,81)
    \Line(288,81)(256,113)
    \Line(288,81)(272,129)
    \Line(368,1)(400,-31)
    \Line(368,1)(384,-47)
    \Line(368,1)(416,-15)
    \Line(368,81)(400,113)
    \Line(288,1)(256,-31)
    \Line[dash,dashsize=10](328,113)(328,-31)
    \Line[dash,dashsize=10](256,41)(400,41)
    \Text(326,-45)[lb]{\Large{\Black{$s$}}}
    \Text(411,36)[lb]{\Large{\Black{$t$}}}
    \Text(245,-41)[lb]{\Large{\Black{$i$}}}
    \Text(410,117)[lb]{\Large{\Black{$j$}}}
    \Text(405,-52)[lb]{\Large{\Black{$Q$}}}
    \Text(235,124)[lb]{\Large{\Black{$P$}}}
  \end{picture}
}
}
\end{center}
\caption{\it A generic two-mass easy box with massless corners $i$ and $j$ corresponding to the end edges of the internal gluon in the Wilson loop picture.}
\label{generic-2me}
\end{figure}
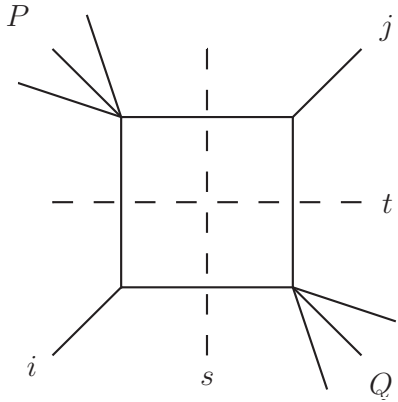\noindent
We must sum over boxes to produce the full amplitude. This yields large cancellations, particularly between non-degenerate boxes in which $i$ and $j$ are separated by at least two intermediate particles. In Appendix \ref{symbology} we show that the $n$-point one-loop MHV amplitude may be written at $\mathcal{O}(\epsilon^0)$ as
\begin{equation}
\begin{aligned}
\label{alternative-mhv-form}
\frac {A_n^\lp}{  A_n^\tr} &= \frac{1}{2}\sum_i \sum_{j \not \in \{i-2,i-1,i,i+1,i+2\} }\left(-\textrm{Li}_2(1 - u_{ij})+\log x_{ij}^2 \log u_{ij} \right) \\
&+\sum_i \log(x_{i i-2}^2)\log\left(\frac{x_{i+1i-2}^2}{x_{i+1 i-1}^2 \sqrt{x_{i i-2}^2}  }\right) \, ,
\end{aligned}
\end{equation}
where $u_{ij}$ denotes the dual conformal invariant cross-ratio,
\begin{equation}
u_{ij}=\frac{x_{ij+1}^2x_{i+1j}^2}{x_{ij}^2x_{i+1j+1}^2} \, ,
\end{equation}
and the square root arises from the infrared divergent pieces (\ref{ir-div-pieces}). The formula (\ref{alternative-mhv-form}) phrases the MHV one-loop amplitude in a form which illuminates dual conformal properties. For example, it is particularly easy to verify that (\ref{dual-conformal-anomaly}) holds. Presently, we shall see that this expression is especially convenient for extracting subleading soft behaviour. We anticipate that this form of the amplitude may find useful further applications. \\\\
We should point out that our description (\ref{alternative-mhv-form}) is not entirely new; partial results of a similar flavour exist in the literature, confirming our observations. The double sum,
\begin{equation}\label{generic-finite-sum}
\frac{1}{2}\sum_i \sum_{j \not \in \{i-2,i-1,i,i+1,i+2\} }\left(-\textrm{Li}_2(1 - u_{ij})+\log x_{ij}^2 \log u_{ij} \right) \, ,
\end{equation}
emerges from considering only non-degenerate two-mass easy-boxes, and accords with the results of \cite{1102.0062}. The remaining sum,
\begin{equation}\label{edge-terms}
\sum_i \log(x_{i i-2}^2)\log\left(\frac{x_{i+1i-2}^2}{x_{i+1 i-1}^2 \sqrt{x_{i i-2}^2}  }\right)\, ,
\end{equation}
comprises degenerate box and infrared divergent contributions, which were recognised but not calculated in \cite{1105.5606}. \\\\
Note that although (\ref{generic-finite-sum}) and (\ref{edge-terms}) look symmetric under reversing the polygon labelling, this is not the case. On careful inspection, we see that this asymmetry is a consequence of our particular choices in arriving at (\ref{combination-of-symbol-terms}). Of course, the symmetry is restored when generic terms and edge cases are summed. 
\\\\
\noindent We first compute the subleading soft behaviour of the generic terms (\ref{generic-finite-sum}). Without loss of generality we consider only those terms in which $i\in \{n-1,n,1\}$. It is convenient to use momentum twistor variables \cite{0905.1473}; see Appendix \ref{momentum-twistors} for a brief review of twistor definitions and identities. In such variables the soft limit may be expressed as  \cite{1406.5155}
\begin{equation}\label{soft-twistor}
Z_n\to \alpha Z_1 +\beta Z_{n-1} + \delta Z_n \, .
\end{equation}
To be precise, (\ref{soft-twistor}) corresponds to the antiholomorphic soft limit with the symmetric elimination $(|n-1],|1])$ as proved in \cite{1406.5155}. Of course box functions transform with zero weight under little group scaling. Hence we may freely switch between holomorphic and antiholomorphic soft limits without affecting our results. For consistency with the Cachazo-Strominger subsitution (\ref{mom-cons-prescription}) we must choose
\begin{equation}\label{alphabetadet}
\alpha=\frac{\l n-1 \ n \r}{\l n-1 \ 1 \r}(1- \delta)\quad\textrm{and}\quad \beta=\frac{\l n \ 1 \r}{\l n-1 \ 1 \r}(1- \delta) \, .
\end{equation}
We have calculated the soft expansion of (\ref{generic-finite-sum}) by employing formulae in Appendix \ref{momentum-twistors}. There are significant telescopic cancellations, yielding bulk terms,
\begin{equation}
\begin{aligned}
&
\frac{1}{\alpha}\sum_{j=4}^{n-4}\log\left(\frac{y_{n-1j}^2}{y_{1j}^2}\right)\frac{\l n-2 \ n-1 \ j-1 \ j \r\l n-2 \ n-1 \ n \ 1 \r}{\l n-2 \ n-1 \ 1 \ j-1 \r \l n-2 \ n-1 \ 1 \ j \r}
 \\
&
+\frac{1}{\beta}\sum_{j=5}^{n-3}\log\left(\frac{y_{2j}^2}{y_{1j}^2}\right)\frac{\l n-1 \ n \ 1 \ 2 \r\l j-1 \ j \ 1 \ 2 \r}{\l n-1 \ 1 \ 2 \ j \r\l n-1 \ 1 \ 2 \ j-1 \r}\, ,
\label{alphabeta-sum}
\end{aligned}
\end{equation}
and boundary contributions,
\begin{equation}
\begin{aligned}
&-\frac{(n-1 \ 1 \ 2 )(n \ 1)}{(n-1 \ 1)(1 \ 2 )}\left[\log\left(\frac{(n-1 \ 1 \ 2 )(n \ 1)}{(n-1 \ 1)(1 \ 2 )}\right)-1\right]+\left(\frac{(n \ 1 ) + (n \ 2 )}{(1 \ 2 )}\right)\left[\vphantom{\frac{A}{B}}\log(-(n \ 1 ))-1\right]
 \\
&
+\frac{(n-1 \ 1 \ 2 )\l n \ 1 \r\left([n \ 1 ]\l 1 \ 3 \r+[n \ 2 ]\l 2 \ 3 \r\right)}{(1 \ 2 )\l n-1 \ 1 \r\left([n-1 \ 1 ]\l 1 \ 3 \r+[n-1 \ 2 ]\l 2 \ 3 \r\right)}\log\left(\frac{(1 \ 2)}{(n-1 \ 1 \ 2 )}\right)
 \\
&
-\frac{\l n-1 \ n \r [n \ 1 ] \l 3 \ 4 \r [2 \ 3]}{[1 \ 2 ]\l n-1 \ 1 \r ([1 \ 2]\l 2 \ 4 \r+ [1 \ 3 ] \l 3 \ 4 \r )} \log\left(\frac{(2 \ 3 )}{(1 \ 2 \ 3 )}\right)
 \\
&
+\left(\frac{(n \ 1 ) + (n \ 2 )}{(1 \ 2 )}-\frac{\l n-1 \ n \r[n \ 2]}{\l n-1 \ 1 \r[1 \ 2 ]}\right)\log(-(n-1 \ 1))+( i \leftrightarrow n-i ) \, .
\end{aligned}
\end{equation}
We have verified this result using box functions and Mathematica numerics in the case $n = 7$. The computations are available as a package, documented in Appendix \ref{package}.
\\\\
Observe that the $\log \delta $ terms take a universal and simple form in the MHV sector for all $n$, namely
\begin{equation}\label{allmhvlogterms}
\left(\frac{(n \ 1) + (n \ 2)}{(1 \ 2 )}-\frac{s_{n-1,1,2}(n \ 1 )}{(n-1 \ 1)(1 \ 2)}\right)\log(-(n \ 1))+( i \leftrightarrow n-i ) \, .
\end{equation}
These structures were already visible at six points in (\ref{logdel6}). Note also that the purely rational terms have a similar universal behaviour. 
\\\\
We must now check that (\ref{allmhvlogterms}) is consistent with the coefficients of $Z_0$ predicted in (\ref{predictedlogsoft}). To see this, we first recall the ansatz (\ref{loopsoftthm}), implicitly eliminating $(|n-1],|1])$ as discussed in Section \ref{summary-verification}.
\begin{equation}\label{loopsoft2}
A_n^\lp \to \left(\frac{1}{\delta^2} S^{(0)\lp}  + \frac{1}{\delta} Z\right) A_{n-1}^\tr \, .
\end{equation}
We focus exclusively on infrared finite $\delta^{-1}\log \delta $ terms. Then the left hand side is given by $A_n^\tr$ times (\ref{allmhvlogterms}). The right-hand side comprises the feed-down term%
\footnote{This is obtained by expanding $S^{(0)\lp}$ to subleading order in $\delta$, with the given momentum conservation prescription. For a full treatment of such subtleties, see Section \ref{tree-soft-sec} and in particular equation (\ref{feed-down}).}
\begin{equation}
\label{feed-down-eg}
-\frac{(n \ 1)+(n-1 \ n )}{(n-1 \ 1 )}S^{(0)}\log(-(n \ 1 )) + (i \leftrightarrow n-i) \, ,
\end{equation}
and the soft anomaly
\begin{equation}\label{z0ansatz}
Z_0|_{\log\delta} = \left(\frac{( n-1 \ n )}{(n-1 \ 1)}+\frac{(n \ 2)}{(1 \ 2 )}-\frac{(n-1 \ 2)(n \ 1 )}{(1 \ 2)(n-1 \ 1)}\right)S^{(0)}\log(-(n \ 1)) + (i \leftrightarrow n-i) \, .
\end{equation}
Summing (\ref{feed-down-eg}) and (\ref{z0ansatz}) yields
\begin{equation}
\left(\frac{(n \ 1 ) + (n \ 2)}{(1 \ 2 )}-\frac{(n \ 1 )}{( 1 \ 2 )}-\frac{( n \ 1 )}{(n-1 \ 1)}-\frac{(n-1 \ 2)(n \ 1 )}{(1 \ 2)(n-1 \ 1)}\right)S^{(0)}\log(-(n \ 1)) + (i \leftrightarrow n-i)  \, ,
\end{equation}
which is identical to \eqref{allmhvlogterms} up to $S^{(0)}$, as expected.

\subsection{NMHV Amplitudes via Ratio Functions}
In the NMHV sector tree-level superamplitudes can be conveniently expressed in terms of dual superconformal $R$-invariants \cite{0807.1095,0808.0491,0808.2475} as
\begin{equation}\label{nmhv-tree}
A_n^{\nmhv,\tr} \ =  \ A_n^{\mhv,\tr}\sum_{j,k}R_{1jk}\, ,
\end{equation}
with $1<j-1$ and $j < k-1$. A general $R$-invariant is most naturally written in terms of momentum supertwistors as \cite{0909.0250}
\begin{equation}
R_{ijk}=\frac{\delta^{(4)}\left(\vphantom{x_1^j}\l j-1 \ j \ k-1 \ k\r\chi_i^A+\textrm{cyclic}\right)}{\l i \ j-1 \ j \ k-1 \r\l j-1 \ j \ k-1 \ k \r\l j \ k-1 \ k \ i \r\l k-1 \ k \ i \ j-1 \r\l k \ i \ j-1 \ j \r}\, ,
\end{equation}
motivating the five-bracket notation,
\begin{equation}
\label{rrr}
R_{ijk}\ :=   \  [i \ j-1 \ j \ k-1 \ k ]\, .
\end{equation}
It is natural to ask whether the $R$-invariants have simple subleading soft behaviour, a study partially undertaken in \cite{1406.5155}. There it was shown that in the supersoft limit,
\begin{equation}\label{supersoft-twistor-limit}
\mathcal{Z}_n \to \alpha \mathcal{Z}_1+\beta \mathcal{Z}_{n-1}+\delta \mathcal{Z}_n \, ,
\end{equation}
the $R$-invariants $R_{1 j k}$ vanish at subleading order. Indeed when $k\neq n$, clearly $R_{1jk}$ is independent of $\delta$, so there is no subleading contribution. For $k= n$, the denominator becomes
\begin{equation}
\delta^2 \l 1 \ j-1 \ j \ n-1 \r\l j-1 \ j \ n-1 \ 1 \r\l j \ n-1 \ n \ 1 \r\l n-1 \ n \ 1 \ j-1 \r\l n-1 \ 1 \ j-1 \ j \r+\mathcal{O}(\delta^3)\, ,
\end{equation}
while the argument of the $\delta$ function is
\begin{equation}
\alpha\l j-1 \ j \ n-1 \ 1 \r\chi_1 + \beta\l n-1 \ 1 \ j-1 \ j \r\chi_{n-1} + \l 1 \ j-1 \ j \ n-1 \r (\alpha \chi_1 + \beta \chi_{n-1}) + \mathcal{O}(\delta)\, .\label{del-fn-arg}
\end{equation}
Notice that the leading term in (\ref{del-fn-arg}) exactly vanishes, hence the leading contribution of the numerator is $\mathcal{O}(\delta^4)$. Therefore $R_{1jn}$ certainly vanishes at subleading order, as claimed. 
\\\\
Recall that for appropriate $\alpha$ and $\beta$ the momentum conservation prescription associated with (\ref{supersoft-twistor-limit}) is exactly the symmetric elimination of $(|n-1],|1])$ (see Appendix E for details). With this prescription the subleading term for tree amplitudes vanishes. Hence we conclude that each $R_{1jk}$ individually obeys the amplitude soft theorem.\\\\
At one loop we may write a general planar NMHV amplitude in terms of dual conformal ratio functions $\R$ as
\begin{equation}
\label{dec}
A_n^{\nmhv,\lp}\ = \  A_n^{\mhv,\lp}\R^\tr \, + \, A_n^{\mhv,\tr}\R^\lp \, .
\end{equation}
$\R^\tr$ is the sum of $R$-invariants appearing in (\ref{nmhv-tree}). $\R^\lp$ may be expressed in terms of general $R$-invariants and dual conformal combinations of box integrals called $V$-functions \cite{0807.1095,0808.0491,0905.4379}. 
\\\\
We now investigate the subleading soft behaviour of the $R$-invariants and $V$-functions appearing at one loop for six- and seven-point amplitudes, leaving general results to future work.   More precisely we will focus on  terms of order $\delta \log \delta$ in $\R^\lp$ which, taking into account the $A_n^{\mhv,\tr}$ prefactor lead to terms of order 
$(1 / \delta) \log \delta$. 
%More in detail, the various terms in \eqref{dec} scale with the soft parameter as 
For illustration we outline the soft expansion of the various terms in \eqref{dec},
\beqa
  A_n^{\mhv,\lp} \ &\!\!\!\sim \!\!\!&\ {1\over \delta^2} + {1\over \delta} \log \delta + {1\over \delta} \ , \qquad
 \R^\tr \ \sim \ 1 + \delta^2 \ , 
 \\ \nonumber 
  A_n^{\mhv,\tr}&\!\!\!\sim\!\!\!& {1\over \delta^2} \ , 
  \qquad \qquad \qquad \qquad 
 \R^\lp \sim 1 + \delta \log \delta + \delta
 \ ,
 \eeqa
where we employ a symmetric momentum conservation prescription that eliminates
$|n-1]$ and $|1]$.
The particular behaviour  for $\R^\tr$ was first observed in \cite{1406.5155}.

\subsubsection*{Six-Point Amplitude}
At six points each five-bracket necessarily omits exactly one momentum twistor. This naturally provides a more concise notation by virtue of the cyclic symmetry of five-brackets. For example we write
\begin{equation}
\label{rb}
(2)=[1 \ 3 \ 4 \ 5 \ 6]\, .
\end{equation}
The six-point tree-level ratio function may then be written as\footnote{The equivalent representations are best understood to arise from Grassmannian contour integration \cite{0907.5418}.}
\begin{equation}
\mathcal{R}^{\tr}=(1)+(3)+(5)=(2)+(4)+(6) \, ,
\end{equation}
which in the soft limit takes the form,
\begin{equation}
\mathcal{R}^{\tr}=(6)+\mathcal{O}(\delta^2) \, ,
\end{equation}
noting that $(6) = [12345]$ has no $\delta$ dependence. The six-point one-loop ratio function is explicitly \cite{1111.1704},
\begin{equation}\label{R1lp}
\R^\lp=\frac{1}{2}\left(\vphantom{\frac{A}{B}}\big[(1)+(4)\big]V_3 + \big[(2)+(5)\big]V_1 + \big[(3)+(6)\big]V_2\right) \, ,
\end{equation}
where the dual conformal $V$-functions are naturally expressed in terms of cross-ratios,
\begin{align}
V_1 &= -\log(u_{36})\log(u_{25})+X \, , \\
V_2 &= -\log(u_{36})\log(u_{14})+X \, ,\\
V_3 &= -\log(u_{14})\log(u_{25})+X \, ,\\
X&= \frac{1}{2}\sum_{i=1}^3 \left(\vphantom{\frac{A}{B}} \log(u_{i i+3})\log(u_{i+1 i+4})+\textrm{Li}_2 (1-u_{i i+3}) \right)-2 \zeta_2 \, ,
\end{align}
and indices in $X$ are implicitly modulo six. We know from tree-level reasoning that
\begin{equation}
(2) = \mathcal{O}(\delta^2),\quad (3) = \mathcal{O}(\delta^2),\quad (4) = \mathcal{O}(\delta^2) \, .
\end{equation}
Therefore in the soft limit the ratio function (\ref{R1lp}) becomes
\begin{equation}\label{R1lpsimpler}
\R^\lp=\frac{1}{2}\left(\vphantom{\frac{A}{B}}(6)\sum_{i=1}^2\textrm{Li}_2 (1-u_{i i+3})+[(6)-(1)]\log(u_{36})\log(u_{25})+[(6)-(5)]\log(u_{36})\log(u_{14})\right) +\mathcal{O}(\delta^2) \, .
\end{equation}
From Appendix \ref{momentum-twistors} observe that
\begin{equation}
\log(u_{36}) = \frac{\delta\l 1 \ 3 \ 5 \ 6 \r}{\alpha\beta\l 1 \ 2 \ 3 \ 5 \r\l 1 \ 3 \ 4 \ 5 \r}\left(\beta\l 2 \ 3 \ 4 \ 5 \r-\alpha\l 1 \ 2 \ 3 \ 4 \r\right)+\mathcal{O}(\delta^2)\, .
\end{equation} 
Hence we need only expand the $R$-invariants to leading order, viz.
\begin{gather}
(1) \ =  \ (6)\frac{\alpha \l 1 \ 2 \ 3 \ 4\r}{\alpha \l 1 \ 2 \ 3 \ 4 \r - \beta \l 2 \ 3 \ 4 \ 5 \r}+\mathcal{O}(\delta)\, , \\
(5) \  =  \  (6)\frac{\beta \l 2 \ 3 \ 4 \ 5 \r}{\beta \l 2 \ 3 \ 4 \ 5 \r - \alpha \l 1 \ 2 \ 3 \ 4 \r}+\mathcal{O}(\delta)\, .
\end{gather}
Thus (\ref{R1lpsimpler}) reduces to
\begin{equation}
\R^\lp=\frac{1}{2}(6)\left(\vphantom{\frac{A}{B}}\sum_{i=1}^2\textrm{Li}_2 (1-u_{i i+3})+\frac{\delta\l 1 \ 3 \ 5 \ 6 \r\l 2 \ 3 \ 4 \ 5\r}{\alpha\l 1 \ 2 \ 3 \ 5 \r\l 1 \ 3 \ 4 \ 5 \r}\log(u_{25})+\frac{\delta\l 1 \ 3 \ 5 \ 6 \r\l 1 \ 2 \ 3 \ 4\r}{\beta\l 1 \ 2 \ 3 \ 5 \r\l 1 \ 3 \ 4 \ 5 \r}\log(u_{14})\right) +\mathcal{O}(\delta^2) \, .
\end{equation}
It is instructive to extract the  $\delta \log\delta$ terms, for these have the best hope of universal behaviour. Explicitly we find the contribution,
\begin{equation}\label{6pt-nmhv-res}
\frac{1}{2}\frac{\l n-1 \ 1 \r}{\l n-1 \ n \r}(6)\frac{ \l 2 \ 3 \ 4 \ 5 \r\l 4 \ 5 \ 6 \ 1 \r}{\l 1 \ 2 \ 4 \ 5 \r\l 3 \ 4 \ 5 \ 1 \r}\log(u_{25})+( i \leftrightarrow n-i )\, .
\end{equation}
\subsubsection*{Seven-Point Amplitude}
At seven points we employ the formulae of \cite{0905.4379}, namely
\begin{equation}\label{Rlp7}
\R^\lp = \frac{1}{2}\left(\R^\tr V^{\textrm{tot}} + R_{147}V_{147} + R_{157}V_{157}+\textrm{cyclic}\right)\, ,
\end{equation}
where the $V$-functions are defined by
\begin{align}
7V^{\textrm{tot}}&=-\textrm{Li}_2(1-u_{1246}^{-1})+\frac{1}{2}\left[\vphantom{\frac{A}{B}}\textrm{Li}_2(1-u_{14})+\textrm{Li}_2(1-u_{15})\right]-\log(u_{47})\log(u_{26})+\textrm{cyclic}\, ,\\
V_{147}&=\textrm{Li}_2(1-u_{2476})+\textrm{Li}_2(1-u_{2146})+\log(u_{2476})\log(u_{2146})-\zeta_2\, ,\\
V_{157}&=V_{147}+\textrm{Li}_2(1-u_{2745})-\textrm{Li}_2(1-u_{1254})-\log(u_{7145})\log\left(\frac{u_{1256}}{u_{2467}}\right)\, ,
\end{align}
and general cross-ratios are written as 
\begin{equation}
u_{ijkl}=\frac{x_{ik}^2 x_{jl}^2}{x_{il}^2 x_{jk}^2 }=\frac{\l i-1 \ i \ k-1 \ k \r\l j-1 \ j \ l-1 \ l \r}{\l  i-1 \ i \ l-1 \ l \r\l j-1 \ j \ k-1 \ k \r}\, .
\end{equation}
We first examine the soft behaviour of the fourteen $R$-invariants explicitly entering (\ref{Rlp7}). Eight of these have no $\mathcal{O}(\delta)$ term, namely
\begin{gather}
R_{147},\ R_{157},\ R_{261},\ R_{372} \, \sim \, \mathcal{O}({\delta^2}) \, , \\
R_{362},\ R_{524},\ R_{625},\ R_{635} \ \   \textrm{independent of }\delta\, .
\end{gather}
We obtain terms linear in $\delta$ from the remaining six, which are
\begin{equation}
R_{251},\ R_{514},\ R_{473},\ R_{413},\ R_{736}, \ R_{746} \, .
\end{equation}
At six points, we had no need to expand such $R$-invariants, courtesy of convenient behaviour of the $V$-functions. We must ask whether this property continues to hold at seven points. Hence we list the $\delta$ dependence of the relevant $V$-functions,
\begin{equation}
\begin{aligned}
V_{251}\sim \mathcal{O}(\delta)\, ,& \qquad V_{736}\sim \mathcal{O}(\delta) \, ,\\
V_{514}\sim \textrm{nonzero}+\mathcal{O}(\delta)\, ,& \qquad V_{473}\sim \textrm{nonzero}+\mathcal{O}(\delta) \, ,\\
V_{746}-V_{736}\sim \textrm{nonzero}+\mathcal{O}(\delta)\, ,& \qquad V_{413}-V_{473}\sim \mathcal{O}(\delta)\, .
\end{aligned}
\end{equation}
We also note that
\begin{equation}
V_{746}-V_{514}\,   \sim \, \mathcal{O}(\delta)\, .
\end{equation}
Therefore the only non-trivial $R$-invariants we must expand to $\mathcal{O}(\delta)$ are the combinations,
\begin{equation}
R_{413}+R_{473} \ \ \ \textrm{ and } \ \ \ R_{514}+R_{746}\, .
\end{equation}
Remarkably, through an intricate series of twistor bracket identities, both of these combinations have zero subleading soft dependence. Thus it only remains to expand the relevant $V$-functions explicitly. Henceforth we shall only look for  $\delta\log\delta$ terms, these being the best candidates for universal behaviour. 
\\\\
It is convenient to express the $V$-functions only in terms of our earlier $u_{ij}$ cross-ratios, defined by
\begin{equation}
u_{ij}=\frac{x_{ij+1}^2x_{i+1j}^2}{x_{ij}^2x_{i+1j+1}^2} \, .
\end{equation}
whose soft expansions are collected in Appendix \ref{momentum-twistors}. Observe that
\begin{equation}
u_{ii+1jj+2}^{-1}=u_{ij}u_{ij+1}\, ,
\end{equation}
and we trivially have relations, 
\begin{equation}
u_{ijkl}=u_{ijlk}^{-1}=u_{klij}\, .
\end{equation}
Thus we may write
\begin{align}
7V^{\textrm{tot}}&=-\textrm{Li}_2(1-u_{14}u_{15})+\frac{1}{2}\left[\vphantom{\frac{A}{B}}\textrm{Li}_2(1-u_{14})+\textrm{Li}_2(1-u_{15})\right]-\log(u_{47})\log(u_{26})+\textrm{cyclic}\, ,\\
V_{147}&=\textrm{Li}_2(1-u_{62}u_{63})+\textrm{Li}_2(1-u_{14}u_{15})+\log(u_{62}u_{63})\log(u_{14}u_{15})-\zeta_2\, ,\\
V_{157}&=V_{147}+\textrm{Li}_2(1-u_{47}u_{41})-\textrm{Li}_2(1-u_{14})+\log(u_{74})\log\left(\frac{u_{62}u_{63}}{u_{15}}\right)\, .
\end{align}
We only obtain $\log\delta$ terms from the invariants $u_{15}$ and $u_{26}$. By relabelling symmetry it suffices to determine only the $\log( u_{26})$ terms. The leading behaviour of the $R$-invariants involved is 
\begin{gather}
\!\!\!\!R_{473}+R_{413} \to (5),\quad R_{625} \to (3),\quad R_{635} \to (1)\, ,\\
R_{413} \to \frac{\beta \l 2 \ 3 \ 4 \ 6 \r}{\beta \l 2 \ 3 \ 4 \ 6 \r - \alpha \l 1 \ 2 \ 3 \ 4 \r}(5)\, , \\
R_{251} \to \frac{\beta \l 2 \ 4 \ 5 \ 6 \r}{\beta \l 2 \ 4 \ 5 \ 6 \r - \alpha \l 1 \ 2 \ 4 \ 5 \r } (3)\, .
\end{gather}
On expanding the relevant $V$-functions many terms are produced. Quite unexpectedly, when multiplying by the respective $R$-invariants a highly non-trivial simplification takes place, yielding the expression,
\begin{equation}
\begin{aligned}
&
R^\tr \ \frac{\delta}{\alpha}\left(\frac{\l 3 \ 5 \ 4 \ 6 \r \l 7 \ 6 \ 1 \ 4 \r}{\l 6 \ 1 \ 3 \ 4 \r\l 6 \ 1 \ 4 \ 5 \r}+\frac{\l 3 \ 4 \ 5 \ 6 \r\l 1 \ 2 \ 6 \ 7 \r}{\l 3 \ 4 \ 6 \ 1 \r\l 1 \ 2 \ 5 \ 6 \r}+\frac{\l 2 \ 3 \ 4 \ 5 \r \l 1 \ 2 \ 6 \ 7 \r}{\l 1 \ 2 \ 4 \ 5 \r\l 2 \ 3 \ 6 \ 1 \r}-\frac{\l 1 \ 2 \ 6\ 7 \r\l 2 \ 3 \ 5 \ 6 \r}{\l 1 \ 2 \ 5 \ 6 \r\l 2 \ 3 \ 6 \ 1 \r}\right) \\
&+(5)\frac{\l 2 \ 3 \ 4 \ 6 \r \l 5 \ 6 \ 7 \ 1 \r}{\l 3 \ 4 \ 6 \ 1 \r \l 1 \ 2 \ 5 \ 6 \r}+(3)\frac{\l 2 \ 4 \ 5 \ 6 \r\l 5 \ 6 \ 7 \ 1 \r}{\l 4 \ 5 \ 6 \ 1 \r\l 1 \ 2\ 5 \ 6 \r}\, .
\end{aligned}
\end{equation}
All that remains is to extract the $\log(u_{26})$ pieces from $V^{\textrm{tot}}$. These come from
\begin{equation}\label{evensimpler-logu26coeff}
-\log(u_{47}u_{41})\log(u_{26})-\textrm{Li}_2(1-u_{25}u_{26})-\textrm{Li}_2(1-u_{62}u_{63})+\textrm{Li}_2(1-u_{26})\, ,
\end{equation}
yielding subleading soft terms,
\begin{equation}
-\frac{\delta}{\alpha}\left(\frac{\l 3 \ 5 \ 4 \ 6 \r\l 7 \ 6 \ 1 \ 4 \r}{\l 6 \ 1 \ 3 \ 4\r\l 6 \ 1 \ 4 \ 5 \r}+\frac{\l 2 \ 3 \ 4 \ 5 \r\l 1 \ 2 \ 6 \ 7 \r}{\l 1 \ 2 \ 4 \ 5 \r\l 2 \ 3 \ 6 \ 1 \r}+\frac{\l 3 \ 4 \ 5 \ 6 \r\l 1 \ 2 \ 6 \ 7 \r}{\l 1 \ 2 \ 5 \ 6 \r\l 3 \ 4 \ 6 \ 1 \r}-\frac{\l 1 \ 2 \ 6 \ 7 \r\l 2 \ 3 \ 5 \ 6 \r}{\l 1 \ 2 \ 5 \ 6 \r\l 2 \ 3 \ 6 \ 1 \r}\right)\, .
\end{equation}
Miraculously these terms exactly cancel terms in (\ref{evensimpler-logu26coeff}). Hence we arrive at the final expression for subleading $\log\delta$ contributions,
\begin{equation}\label{nmhv7ptres}
\frac{1}{2}\frac{\l n-1 \ 1 \r}{\l n-1 \ n \r}\left[(5)\frac{\l 2 \ 3 \ 4 \ 6 \r \l 5 \ 6 \ 7 \ 1 \r}{\l 3 \ 4 \ 6 \ 1 \r \l 1 \ 2 \ 5 \ 6 \r}+(3)\frac{\l 2 \ 4 \ 5 \ 6 \r\l 5 \ 6 \ 7 \ 1 \r}{\l 4 \ 5 \ 6 \ 1 \r\l 1 \ 2\ 5 \ 6 \r}\right]\log(u_{26})+ ( i \leftrightarrow n-i )\, .
\end{equation}
Observe that these terms have the same overall structure as we found at six points in equation (\ref{6pt-nmhv-res}). Furthermore, one may immediately perform a consistency check that (\ref{nmhv7ptres}) reduces to (\ref{6pt-nmhv-res}) as we make particles $3$ and $4$ collinear. These terms seem amenable to an $n$-point generalisation, which we leave to future work. \\\\
We note finally that the coefficients in (\ref{nmhv7ptres}) were not predicted in Section \ref{sec:sub-soft-1-lp}; indeed they involve particles other than $\{n-2, n-1, n, 1 , 2\}$ which were considered in deriving (\ref{predictedlogsoft}). It would be interesting to investigate whether the constraint equation (\ref{Z0-constraint}), perhaps supplemented with further physical reasoning, is sufficently powerful to determine the NMHV one-loop subleading soft anomaly in general.

\section{Conclusions}
The (sub)leading soft behaviour of one-loop amplitudes has received much attention recently. However, obtaining general results for the infrared-finite terms of amplitudes in non-abelian gauge theories has proved to be difficult \cite{1405.1015, 1405.1410, 1406.6987,1503.05156}. In this work we computed the subleading soft behaviour of infrared-finite parts of all one-loop MHV amplitudes in planar $\mathcal{N}=4$ SYM, exposing surprising hidden simplicity. Moreover, we determined the subleading soft contributions of NMHV one-loop ratio functions at six and seven points, finding evidence that universality holds within but not between helicity sectors.
\\\\
Interestingly, none of the available representations of the one-loop amplitudes makes the subleading soft behaviour manifest. Indeed, in both MHV and NMHV sectors highly non-trivial cancellations occur, which lead to compact formulae for the soft limits. We speculate that amplitudes may admit a recasting in ``soft friendly'' form, perhaps along the lines of \cite{1303.4734}. At higher loops this might yield a subleading soft-improved BDS ansatz
\cite{hep-th/0505205}. 
\\\\
We also used the known one-loop dual superconformal anomaly of amplitudes  to derive  constraints on one-loop soft limits of generic amplitudes. In this respect we introduced the 
soft anomaly $Z$ in \eqref{S1lp}, which we then studied up to and including infrared-finite terms, thus predicting the general form of the $(1/\delta) \log \delta $ terms, see \eqref{predictedlogsoft}. 
This constraint turned out to be  sufficiently powerful to predict the complete MHV soft anomaly, and provided a valuable consistency check in other sectors. It would be fascinating to understand in greater detail how quantum corrections to soft theorems interact with the asymptotic symmetries of $\mathcal{N}=4$ SYM \cite{1308.0589}.
\\\\
Soft theorems lie at a crossroads of theory and phenomenology. Consequently several  avenues for further work naturally present themselves. From a mathematical perspective, we might hope to derive our results from recent ambitwistor formulae \cite{1507.00321,1406.1462}. 
The subleading soft anomaly may also find application in bootstraps for amplitudes \cite{0907.5418, 1108.5385, 1204.4841} and effective field theories \cite{1412.4095}. Extensions to double soft emission and supergravity theories are obvious goals, particularly in light of the subtle connections with spontaneously broken symmetry \cite{0808.1446}.
\\\\ 
Moving towards experiment, our results could provide a stepping stone towards improved soft-gluon approximation for QCD amplitudes. Such an objective would require examining further helicity sectors and theories with less supersymmetry, possibly making contact with \cite{1503.05156}. Finally, it would be profitable to apply our methods to form factors, developing the results of \cite{1506.07551}, and to amplitudes and form factors at higher loops. In particular such computations would be  useful in cases where
the form factors are related to Higgs amplitudes appearing in effective field theory
approaches (see for example \cite{hep-th/0411092,1112.3554,1201.4170}).

\section*{Acknowledgements}
It is a pleasure to thank Zvi Bern, Scott Davies, Francesca Day, Joseph Hayling, Zac Kenton, Martyna Kostacinska, Lorenzo Magnea, James McGrane, Brenda Penante, Jan Plefka and Chris White for useful discussions.  All figures in this paper were drawn with the aid of JaxoDraw \cite{hep-ph/0309015}. This work was supported by the Science and Technology Facilities Council Consolidated Grant ST/L000415/1 \textit{String theory, gauge theory \& duality}.
\appendix
\section{Spinor Conventions}\label{spinor-conventions}
Our index convention for spinors is
\begin{equation}
\lambda^\alpha_i=|i\r^\alpha \, ,
\qquad 
\tilde\lambda_{i\ad}=|i]_\ad \, ,
\qquad 
\lambda_{i\alpha}=\l i|_\alpha =\epsilon_{\alpha \beta}|i\r^\beta \, ,
\qquad
\tilde\lambda^\ad_i=[i|^\ad =\epsilon^{\ad\dot\beta}|i]_{\dot\beta} \, ,
\end{equation}
with $\sigma$ matrices chosen such that
\begin{equation}
(i \ j)=\l i \ j \r[j \ i] \, ,
\end{equation}
where $(i \ j )=2\left(p_i\cdot p_j\right)$. Spinor differentiation is defined to obey
\begin{align}
\frac{\partial}{\partial |i\r^\alpha}|i\r^\beta &= \delta_\alpha^\beta\, ,\notag\\
\frac{\partial}{\partial |i]_\ad} |i]_{\dot\beta} &=\delta^\ad_{\dot\beta} \, .
\end{align}
Note that the helicity operator for particle $i$ takes the form,
\begin{equation}\label{helicity-op}
h_i = -\frac{1}{2}\left(|i\r\cdot\frac{\partial}{\partial|i\r}-|i]\cdot\frac{\partial}{\partial |i]}-\eta_i^A\frac{\partial}{\partial \eta_i^A}\right)\, ,
\end{equation}
and acts as the identity on superamplitudes, since each superparticle has helicity $1$ by construction. This is of  importance in Section \ref{sub-soft-tree}.

\section{Action of the Dual Conformal Boost Generator}\label{useful-dual-conf-formulae}
We collect various formulae outlining the action of the dual conformal boost generator on spinors and multiparticle invariants used in Sections \ref{main} and \ref{verification}. To adapt the formulae to $\left(K_{\alpha\ad}\right)_{\mathcal{O}(\delta^0)}$ replace all $\sum_j$ by $\sum_j'$. 
%\red The correctness of all statements relies heavily upon Schouten identities. \black 
\\\\
Suppose $a<b$ cyclically in $\{3,4,\dots 2\}$. Then we have
\begin{align}
-K(\l a \ b \r)&=\sum_{j=3}^{a-1}|j]\l j|\l a \ b \r+\sum_{j=a+1}^{b-1}|j]\l b| \l a \ j \r\, ,\\
-K([a \ b ])&=\sum_{j=3}^b |j]\l j|[a \ b]+\sum_{j=a+1}^{b-1}|a]\l j| [b \ j ] \, ,\\
-K((a \ b ))&=2\sum_{j=3}^{a-1}|j]\l j| (a \ b)+\sum_{j=a}^b |j]\l j| (a \ b) + \sum_{j=a+1}^{b-1}|j]\l b| \l a \ j\r [b \ a]-\sum_{j=a+1}^{b-1}|a]\l j| [b \ j]\l a \ b\r \, .
\end{align}
In particular if $a$ and $b$ are adjacent then
\begin{align}
-K(\l a \ b \r)&=\sum_{j=3}^{a-1}|j]\l j|\l a \ b  \r \, ,\\
-K([a \ b ])&=\sum_{j=3}^b |j]\l j|[a \ b] \, ,\\
-K((a \ b ))&=2\sum_{j=3}^{a-1}|j]\l j| (a \ b)+\sum_{j=a,b} |j]\l j| (a \ b)  \, .
\end{align}
The following corollaries are of particular use in Section \ref{sec:sub-soft-1-lp}.
\begin{align}
\Big[\big(\K_{\a \ad}\big)_{\mathcal{O}(\delta^0)}-\sum_{j=3}^{n-1}|j]\l j|\Big]\Big(\frac{[2 \ n]}{[2 \ 1] \l 1 \ n \r}\Big)&\ = \ -\frac{|n]\l 1|}{\l n \ 1 \r}\, ,\label{usefulKspartIfirst}\\
\Big[\big(\K_{\a \ad}\big)_{\mathcal{O}(\delta^0)}-\sum_{j=3}^{n-1}|j]\l j|\Big]\Big(\frac{[n-2 \ n ]}{[n-2 \ n-1] \l n-1 \ n \r}\Big)&\ = \ -\frac{|n]\l n-1|}{\l n-1 \ n \r}\, ,\\
\Big[\big(\K_{\a \ad}\big)_{\mathcal{O}(\delta^0)}-\sum_{j=3}^{n-1}|j]\l j|\Big]\Big(\frac{[1 \ n ]}{[1 \ n-1]\l n-1 \ n \r}\Big)&\ = \ -|n-1]\l n-1 | \frac{[n \ 1 ]}{\l n-1 \ n \r[n-1 \ 1 ]}\, ,\\
\big[\big(\K_{\a \ad}\big)_{\mathcal{O}(\delta^0)}-\sum_{j=3}^{n-1}|j]\l j|\Big]\Big(\frac{[n-1 \ n ]}{[n-1 \ 1 ]\l 1 \ n \r}\Big)&\ =\ -|1]\l 1| \frac{[n-1 \ n]}{[n-1 \ 1]\l n\ 1 \r}\, ,\label{usefulKspartIlast}\\
\big(\K_{\a \ad}\big)_{\mathcal{O}(\delta^0)}\Big(\frac{(n \ 1 )}{(n-1 \ 1 )}\Big)&\ =\  -|n-1]\l n-1|\frac{(n \ 1)}{(n-1 \ 1 )}\, ,\label{usefulKspartIIfirst}\\ 
\big(\K_{\a \ad}\big)_{\mathcal{O}(\delta^0)}\Big(\frac{(n-2 \ n)}{(n-2 \ n-1 )}-\frac{(n- 2 \ 1)(n-1 \ n )}{(n-2 \ n-1)(n-1 \ 1)}\Big)&\ =\ -|n]\l n| + |1] \l 1| \frac{( n-1 \ n )}{( n-1 \ 1 )} \, .\label{usefulKspartIIlast}
\end{align}
\section{\texttt{SubSoft.m} Package Documentation}\label{package}
\texttt{SubSoft.m} is a Mathematica package for the automated calculation and verification of subleading soft theorems. A separate Mathematica file contains sample calculations, pertinent to our results in Section \ref{verification}. The package extends Bourjaily's \texttt{bcfw.m} \cite{1011.2447}. The relevant Mathematica files are included with the submission of this posting on the \texttt{arXiv}.\footnote{From the abstract page, follow the link to download ``other formats'' and unzip the resulting tarball. The files \texttt{SubSoft.m} and \texttt{bcfw.m} are required, and \texttt{SubSoft Examples.nb} is an optional walkthrough.}
\subsubsection*{Setup}
First ensure that both \texttt{SubSoft.m} and \texttt{bcfw.m} are saved to the same directory as the notebook you are writing. To initialize the package, simply call
\begin{center}
\includegraphics[width=\textwidth]{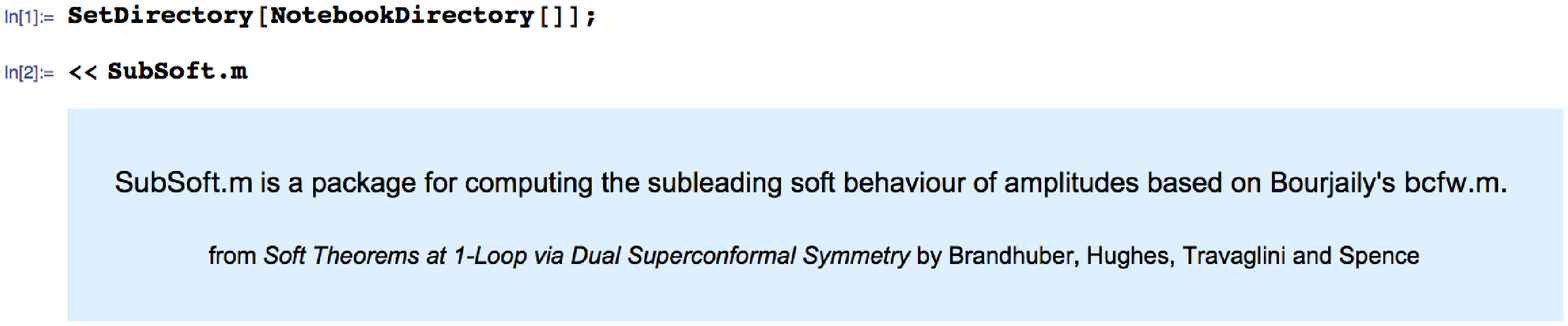}
\end{center}
\subsubsection*{Glossary}
We collect descriptions of the most important expressions. The definitions of related expressions may be inferred, or determined by direct inspection of the source code.
\\
\begin{tabularx}{\textwidth}{p{6cm} p{2cm} X }
\textbf {Expression} & \textbf {Type} & \textbf  {Description}  \\ \hline \endhead 
\texttt{ab[i,j]} & object & represents the angle bracket $\l i \ j\r$. \\
\texttt{sb[i,j]} & object & represents the square bracket $[i \ j ]$. \\
\texttt{MHVTreeAmplitude[\{i,j\},n]} & function & {\raggedright returns the tree amplitude with helicity configuration $1^+ \cdots i^- \cdots j^- \cdots n^+$}. \\
\texttt{F[i,n]} & function & returns the $i^\textrm{th}$ box function for $n$-point kinematics. \\
\texttt{useRandomKinematics[n]} & function & sets up $n$-point random kinematics for numerical evalulation. \\
\texttt{NEvalute[expr]} & function &numerically evaluates an expression featuring \texttt{ab} and/or \texttt{sb}. \\
\texttt{deltaDependence[n]} & rule & introduces holomorphic $\delta$ dependence for particle $n$. \\
\texttt{momentumConservation}\newline\texttt{WithDelta[n,a,b]} &  rule & performs the substitution (\ref{mom-cons-prescription}) assuming that particle $n$ carries $\delta$ dependence. \\
$\mathtt{VerifySoftTheorem}$\newline$\mathtt{TreeLevel[A_n, 
A_{n-1}, n, {a, b}]}$ & function & verifies the tree-level subleading soft theorem for given $A_n$, $A_{n-1}$ with the elimination of $|a]$ and $|b]$. \\
$\mathtt{VerifySoftTheorem1LoopIRFinite}$\newline$\mathtt{Term[Alp_n, Alp_{n-1}, 
 Atr_{n-1}, n, {i, j}]}$ & function & verifies the one-loop subleading soft term at finite order in $\epsilon$ for given $A_n^\lp$, $A_{n-1}^\lp$, $A_{n-1}^\tr$ with the elimination of $|a]$ and $|b]$.\\
 \texttt{Z0[n]} & function & returns the predicted infrared-finite subleading soft anomaly, by default defined for MHV amplitudes with $n=5,6,7$.
\end{tabularx}\noindent
Strictly speaking, the verification functions compute the difference between subleading terms on the LHS and RHS of (\ref{tree-soft-theorem-bnd-integrated}) or (\ref{loopsoftthm}) at tree level or one loop respectively. Hence the subleading soft theorems are verified if the resulting quantity is within machine precision of zero.
\section{Generic Two-Mass Easy Box Cancellations via Symbology}\label{symbology}
The finite part of a two-mass easy box may compactly be defined as \cite{hep-th/0407214}
\begin{equation}
F^{2\textrm{me,fin}}(P,Q,s,t)=\textrm{Li}_2(1-aP^2)+\textrm{Li}_2(1-aQ^2)-\textrm{Li}_2(1-as)-\textrm{Li}_2(1-at)\, ,
\end{equation}
where 
\begin{equation}
a= \frac{P^2+Q^2-s-t}{P^2Q^2 -st}\, .
\end{equation}
It is convenient to write the momentum invariants in terms of differences of dual momenta,
\begin{equation}
P^2=x_{i+1j}^2\, , \quad Q^2 = x_{ij+1}^2\, , \quad s=x_{ij}^2 \, ,\quad t=x_{i+1j+1}^2\, .
\end{equation}
We may derive a more compact form for the sum over generic boxes by introducing the symbol \cite{1006.5703}. This is a map taking transcendental functions to tensor products of their rational arguments. In particular,
\begin{align}
\sym\left[\log (R_a)\log (R_b)\right]&\ = \ R_a\tensor R_b + R_b \tensor R_a\, ,\\
\sym\left[\textrm{Li}_2(1-R_a)\right]&\ = \ -R_a\tensor (1-R_a)\, .
\end{align}
The target space is defined modulo the identifications,
\begin{gather}
R_a R_b \tensor R_c R_d = R_a \tensor R_c + R_b \tensor R_c + R_a \tensor R_d + R_b \tensor R_d \, ,\\
\textrm{constant}\tensor R_a = R_a \tensor \textrm{constant} = 0 \, ,\\
R_a \tensor (R_b)^{-1}= (R_a)^{-1}\tensor R_b = - R_a \tensor R_b\, .
\end{gather}
We evaluate the symbols,
\begin{equation}
\begin{aligned}
\sym\left[\textrm{Li}_2(1-aP^2)\right]&=a\tensor (P^2Q^2 - s t)+P^2 \tensor (P^2 Q^2 - s t)  \\ &- a  \tensor (s-P^2) (P^2 - t) - P^2 \tensor (s-P^2) (P^2 - t) \, ,
 \\
\sym\left[\textrm{Li}_2(1-aQ^2)\right]&=a\tensor (P^2Q^2 - s t)+Q^2 \tensor (P^2 Q^2 - s t)  \\
&- a  \tensor (s-Q^2) (Q^2 - t) - Q^2 \tensor (s-Q^2) (Q^2 - t) \, ,
 \\
\sym\left[\textrm{Li}_2(1-as)\right]&=a\tensor (P^2Q^2 - s t)+s \tensor (P^2 Q^2 - s t)  \\&- a \tensor (P^2-s) (Q^2 - s) - s \tensor (P^2-s) (Q^2 - s) \, ,
 \\
\sym\left[\textrm{Li}_2(1-as)\right]&= a\tensor (P^2Q^2 - s t)+t \tensor (P^2 Q^2 - s t)  \\&- a \tensor (P^2-t) (Q^2 - t) - t \tensor (P^2-t) (Q^2 - t) \, .
\end{aligned}
\end{equation}
The first and third terms in each symbol cancel in the sum defining the symbol of $F^{2\textrm{me},\textrm{fin}}$. The second terms combine to yield
\begin{equation}\label{main-contrib}
\frac{P^2Q^2}{s t}\tensor(P^2Q^2 - st)=\frac{x_{i j+1}^2 x_{i+1 j}^2}{x_{ij}^2 x_{i+1 j+1}^2}\tensor (x_{i+1j}^2 x_{i j+1}^2 - x_{ij}^2 x_{i+1 j+1}^2) \, .
\end{equation}
It is convenient to write the fourth terms in dual variables,
\begin{align}
P^2 \tensor (s-P^2) (P^2 - t) &= x_{i+1 j }^2 \tensor (x_{ij}^2 - x_{i+1 j}^2) + x_{i+1 j }^2 \tensor (x_{i+1 j}^2 - x_{i+1 j+1}^2)\label{symbP2} \, ,\\
Q^2 \tensor (s-Q^2) (Q^2 - t) &= x_{i j+1}^2 \tensor(x_{ij}^2 - x_{ij+1}^2)+x_{ij+1}^2 \tensor (x_{ij+1}^2 - x_{i+1j+1}^2) \, ,\\
s \tensor (P^2-s) (Q^2 - s) &= x_{ij}^2\tensor (x_{i+1j}^2 - x_{ij}^2) + x_{ij}^2 \tensor (x_{i j+1}^2 - x_{ij}^2) \, ,\\
t \tensor (P^2-t) (Q^2 - t) &= x_{i+1 j+1}^2 \tensor(x_{i+1j}^2 - x_{i+1j+1}^2) + x_{i+1 j+1}^2 \tensor (x_{i j+1}^2 - x_{i+1j+1}^2)\, .\label{symbt}
\end{align} 
\iffalse
In the sum defining $\sym \left[F^{2\textrm{me},\textrm{fin}}\right]$ we may pair these terms as follows,
\beqa
&&x_{i+1 j+1}^2 \tensor (x_{i j+1}^2 - x_{i+1j+1}^2) - x_{i+1 j }^2 \tensor (x_{ij}^2 - x_{i+1 j}^2)\, ,  \\
&&x_{ij}^2 \tensor (x_{i j+1}^2 - x_{ij}^2) - x_{i+1 j }^2 \tensor (x_{i+1 j}^2 - x_{i+1 j+1}^2)\, , \\
&&x_{i+1 j+1}^2 \tensor(x_{i+1j}^2 - x_{i+1j+1}^2) - x_{i j+1}^2 \tensor(x_{ij}^2 - x_{ij+1}^2)\, ,  \\
&&x_{ij}^2\tensor (x_{i+1j}^2 - x_{ij}^2) - x_{ij+1}^2 \tensor (x_{ij+1}^2 - x_{i+1j+1}^2) \, .
\eeqa
\fi
To produce the complete finite part of the amplitude we must sum over all distinct boxes. This corresponds to summing over all $i$ and $j$ not adjacent and dividing by a factor of $2$. We now apply this procedure to the symbols (\ref{main-contrib})--(\ref{symbt}) to exhibit hidden cancellations.
\\\\
Consider for fixed $i$ the telescoping sum,
\begin{equation}
\sum_{j\not\in \{i-1,i,i+1\}}A_{i j+1} - A_{ij} = A_{i i-1} - A_{i i+2} \, .
\end{equation}
We may employ this formula to find the contribution of (\ref{symbP2})--(\ref{symbt}) to the full symbol. The resulting term is
\begin{gather}\label{edge1}
\sum_i \sym \left[\log^2 (x_{i i-2}^2) \right] \, .
\end{gather}
We now massage (\ref{main-contrib}) into a form we can integrate, writing
\begin{equation}\label{Li2-origin}
\frac{x_{i j+1}^2 x_{i+1 j}^2}{x_{ij}^2 x_{i+1 j+1}^2}\tensor (x_{i+1j}^2 x_{i j+1}^2 - x_{ij}^2 x_{i+1 j+1}^2)\ =  \  u_{ij}\tensor (1 - u_{ij})+ u_{ij}\tensor x_{ij}^2 x_{i+1 j+1}^2 \, .
\end{equation}
We immediately identify the first term as the symbol of $-\textrm{Li}_2 (1-u_{ij})$. The second term expands to give neatly paired contributions,
\begin{equation}
\begin{aligned}
&x_{i j+1}^2 \tensor x_{ij}^2  + x_{i+1 j }^2 \tensor x_{i+1 j+1}^2 
+x_{i j+1}^2 \tensor x_{i+1 j+1}^2 + x_{i+1 j }^2 \tensor x_{ij}^2\\
&-x_{ij}^2 \tensor x_{i j}^2 - x_{i+1 j+1}^2 \tensor x_{i+1 j+1}^2 
-x_{ij}^2 \tensor x_{i+1 j+1}^2 - x_{i+1 j+1}^2 \tensor x_{ij }^2 \, .
\end{aligned}
\end{equation}
Performing the sum over non-adjacent $i$ and $j$ we find that
\begin{equation}
\begin{aligned}
\label{combination-of-symbol-terms}
&
\sum_i \sum_{j \not \in \{i-2,i-1,i,i+1\}} \sym\left[\log (x_{ij}^2)\log(x_{i j+1}^2)\right]+\sum_i \sum_{j\not \in \{i-1, i, i+1, i+2\}} \sym \left[\log(x_{ij}^2)\log(x_{i+1 j}^2)\right]
 \\
&-\sum_{i}\sum_{j\not \in \{i-1,i,i+1\}}\sym\left[ \log(x_{ij}^2)\log(x_{i+1 j+1}^2)\right]-\sum_{i}\sum_{j\not \in \{i-1,i,i+1\}}\sym \left[\log (x_{ij}^2) \log(x_{ij}^2)\right] \, .
\end{aligned}
\end{equation}
Combining the terms (\ref{edge1}) and (\ref{combination-of-symbol-terms}), integrating the symbol and dividing by $2$ yields
\begin{equation}\label{mainplusedge2}
\sum_i \left[\frac{1}{2}\sum_{j\not \in \{i-2, i-1, i, i+1, i+2\}}\log(x_{ij}^2)\log(u_{ij})+\log(x_{i i-2}^2)\log\left(\frac{x_{i+1i-2}^2}{ x_{i+1 i-1}^2 }\right)\right] \, .
\end{equation}
We finally split our expression for the finite part of the amplitude into generic terms,
\begin{equation}
\frac{1}{2}\sum_i \sum_{j \not \in \{i-2,i-1,i,i+1,i+2\} }\left(-\textrm{Li}_2(1 - u_{ij}) +\log x_{ij}^2 \log u_{ij} \right) \, ,
\end{equation}
and edge cases,
\begin{equation}\label{edge-almost-there}
\sum_i \log(x_{i i-2}^2)\log\left(\frac{x_{i+1i-2}^2}{x_{i+1 i-1}^2 }\right) \, .
\end{equation}
Note that the edge cases comprise the full finite part of the amplitude at five-point, which we have verified by comparison with \cite{hep-ph/9403226}.

\section{Momentum Twistors: Identities and Soft Limits}\label{momentum-twistors}
\subsubsection*{Identities}
Momentum supertwistors form the fundamental representation of dual superconformal symmetry, and are natural variables in which to express amplitudes in $\mathcal{N}=4$ SYM \cite{0905.1473}. We define 
\begin{equation}\label{twistor-def}
{\cal{Z}}_i := (Z_i^I; \chi^A_i) = (\lambda_i^\alpha, \, \mu_i^{\dot\alpha}; \,
\chi^A_i) \, ,
\end{equation}
with  the incidence relations 
\beq
\mu_i^{\dot\alpha}=x_i^{\dot\alpha\beta}\lambda_{i \beta} \, , \ \ 
\chi^A_i=\theta_i^{\beta A} \lambda_{i \beta} \, .
\eeq
Here  $Z_i^I$ denotes the bosonic part of the supertwistor,  with
$I = ({\alpha,\dot\alpha})$. This definition yields a twistor correspondence \cite{Penrose:1967wn} relating points and straight lines. More explicitly, 
\begin{align}
\textrm{dual momentum space}\quad &\longleftrightarrow\quad \textrm{momentum twistor space} \, ,\\
\textrm{point } x_i \quad &\longleftrightarrow\quad \textrm{line through } Z_{i-1} \textrm{ and } Z_i \, ,\\
\textrm{line through } x_i \textrm{ and } x_{i+1} \quad &\longleftrightarrow\quad \textrm{point } Z_i \, .
\end{align}
Importantly the on-shell variables $\tilde{\lambda}_i$ and $\eta_i$ can be obtained from
${\cal{Z}}_i$ using the relations (see for example \cite{0909.0250,Adamo:2011pv})
\beqa\label{twistorformulas}
\tilde{\lambda}_i & = &
\frac{\mu_{i-1} \langle i\ i+1 \rangle + \mu_i \langle i+1\ i-1 \rangle
+ \mu_{i+1} \langle i-1\ i \rangle}{\langle i-1\ i\rangle \langle i\ i+1 \rangle} \ ,
\nonumber\\
\eta_i & = &
\frac{\chi_{i-1} \langle i\  i+1 \rangle + \chi_i \langle i+1\  i-1 \rangle
+ \chi_{i+1} \langle i-1\  i \rangle}{\langle i-1\  i\rangle \langle i\  i+1 \rangle} \ .
\eeqa
The canonical bosonic dual conformal invariant quantity is the four-bracket,
\begin{equation}
\l i \ j \ k \ l\r = \epsilon_{IJKL}Z_i^IZ_j^J Z_k^K Z_l^L  \, ,
\end{equation}
while the definition of the holomophic spinor bracket requires the use of the
infinity twistor ${\cal{I}}$ such that
\begin{equation}
\langle i\ j \rangle = \epsilon_{\alpha\beta} \lambda_i^\alpha \lambda_j^\beta
= \epsilon_{IJKL} Z^I_i Z^J_j {\cal{I}}^{KL} \ .
\end{equation}
These objects obey various identities, which we employ  in Section \ref{mhv-twistors}. From the definition (\ref{twistor-def}) one can show for example that
\begin{equation}
\l i \ j-1 \ j \ k \r = \l j-1 \ j \r\l i | x_{ij}x_{jk} |k \r \, .
\end{equation}
The four-brackets obey a  five-term Schouten identity,
\begin{equation}
Z_a\l b\ c \ d \ e \r + \textrm{cyclic} = 0 \, ,
\end{equation}
which yields formulae for computing intersections of projective lines and planes,
\begin{align}
(i\ j) \cap (a\ b \ c) &= Z_i\l j \ a \ b \ c \r - Z_j \l i \ a \ b \ c \r \, , \\
(i\ j\ k) \cap (a\ b\ c) &= Z_i Z_j \l k \ a \ b \ c \r + Z_j Z_k \l i \ a \ b \ c \r + Z_k Z_i \l j \ a \ b \ c \r \, ,
\end{align}
where we have introduced the notation $(a \ b) = Z_a \wedge Z_b$. Finally we have the important relation,
\begin{equation}
\l x \ y \ (i \ j \ k )\cap(a \ b \ c )\r = \l (x \ y )\cap (a \ b \ c) \ i \ j \ k \r \, .
\end{equation}
%valid as a statement in homogeneous coordinates, where scale is important. To verify this, %first observe that it has the correct vanishing behaviour in the case of linear dependence. %Then it suffices to evaluate one non-vanishing example.

\subsubsection*{Soft Limits}
The supersoft limit of an amplitude may be implemented in momentum twistor variables by taking \cite{1406.5155}
\begin{equation}\label{soft-twistor-appendix}
{\cal{Z}}_n\to \alpha {\cal{Z}}_1 +\beta {\cal{Z}}_{n-1} + \delta {\cal{Z}}_n \, .
\end{equation}
For generic $\alpha$ and $\beta$ four spinors gain $\delta$-dependence, namely $|n-1]$, $|n]$, $|n\r$ and $|1]$. In Section \ref{verification} we require the antiholomorphic supersoft limit,
\beq
|n\r \to |n\r \, , \quad
|n] \to \delta |n] \, , \quad
\eta_n \to \delta \eta_n \, , 
\eeq
with the symmetric elimination of $|n-1]$ and $|1]$. By comparing (\ref{twistorformulas}) to (\ref{mom-cons-prescription}) this stipulation forces
\begin{equation}\label{alphabetadet-appendix}
\alpha=\frac{\l n-1 \ n \r}{\l n-1 \ 1 \r}(1-\delta)\quad\textrm{and}\quad \beta=\frac{\l n \ 1 \r}{\l n-1 \ 1 \r}(1-\delta) \, .
\end{equation}
The $\delta$-dependence of $\alpha$ and $\beta$ is present to ensure that $|n\r$ remains fixed.
\subsubsection*{Soft Expansion of Cross-Ratios and $x_{ij}^2$}
The dual conformal cross-ratio $u_{ij}$ may be expressed as a ratio of twistor four-brackets, namely
\begin{equation}
u_{ij}=\frac{\l i-1 \ i \ j \ j+1 \r \l i \ i+1 \ j-1 \ j \r}{\l i-1 \ i \ j-1 \ j \r \l i\ i+1 \ j \ j+1\r} \, .
\end{equation}
To evaluate the soft behaviour of relevant cross-ratios we will require the $\delta$ expansion
of the four-brackets
\begin{align}
\l n \ 1 \ j-1 \ j \r &= \beta \l n-1 \ 1 \ j-1 \ j \r +\delta \l n \ 1 \ j-1 \ j \r\, ,\\
\l n-1 \ n \ j-1 \ j \r &= \alpha \l n-1 \ 1 \ j-1 \ j \r +\delta \l n-1 \ n \ j-1 \ j \r \, .
\end{align}
Using twistor identities we can then derive simple forms for 
\begin{eqnarray}
u_{n-1 j}&=& v_{n-1 j}\left(1-\frac{\delta\l j-1 \ j \ j+1 \ n-1\r\l n-1 \ n \ 1 \ j \r}{\alpha\l n-1 \ 1 \ j-1 \ j \r\l n-1 \ 1 \ j \ j+1\r}\right) \quad \textrm{ for } 3 \leq j \leq n-4\, ,\\
u_{1 j}&=&v_{1 j}\left(1-\frac{\delta\l j-1 \ j \ j+1 \ 1 \r\l n-1 \ n \ 1 \ j\r}{\beta\l n-1 \ 1 \ j-1 \ j \r \l n-1 \ 1 \ j \ j+1 \r}\right) \quad \textrm{ for } 4 \leq j \leq n-3\, ,\\
u_{n j}&=&1+\frac{\delta\l j-1 \ j+1 \ j \ n-1\r \l n \ n-1 \ 1 \ j \r}{\alpha\l n-1 \ 1 \ j-1 \ j \r\l n-1 \ 1 \ j \ j+1 \r}+\frac{\delta\l j-1 \ j+1 \ 1 \ j \r\l n-1 \ n \ 1 \ j \r}{\beta\l n-1 \ 1 \ j-1 \ j \r\l n-1 \ 1 \ j \ j+1\r} \nonumber \\
&&\textrm{ for } 3 \leq j \leq n-3 \, ,
\end{eqnarray}
valid through subleading order in $\delta$, where the $v_{ij}$ are cross-ratios evaluated in $(n{-}1)$-point kinematics. In addition we have special cases of cross-ratios that vanish
in the soft limit
\beqa
u_{1 n-2}&\!\!=\!\!&\frac{\delta\l n \ 1 \ n-2 \ n-1 \r\l 1 \ 2 \ n-3 \ n-2 \r}{\beta \l 1 \ 2 \ n-1 \ n-2 \r\l n-1 \ 1 \ n-3 \ n-2 \r}\left(1-\frac{\delta \l n \ 1 \ n-3 \ n-2 \r }{\beta \l n-1 \ 1 \ n-3 \ n-2 \r}\right)\, ,\\
u_{n-1 2}&\!\!=\!\!&\frac{\delta\l n-2 \ n-1 \ 2 \ 3 \r\l n-1 \ n \ 1 \ 2 \r}{\alpha\l n-2 \ n-1 \ 1 \ 2 \r\l n-1 \ 1 \ 2 \ 3 \r}\left(1-\frac{\delta\l n-1 \ n \ 2 \ 3 \r}{\alpha \l n-1 \ 1 \ 2 \ 3 \r }\right) \, .
\eeqa
Since these appear as arguments of logarithms
we require these quantities through order $\delta^2$ in order to extract terms subleading
in $\delta$. 
The multiparticle invariants $x_{ij}^2$ may be written as a ratio of a four-bracket to two holomorphic spinor brackets,
\begin{equation}
x_{ij}^2 =\frac{\l i-1 \ i \ j-1 \ j \r}{\l i-1 \ i \r \l j-1 \ j\r} \, ,
\end{equation}
which breaks conformal symmetry due to the presence of the infinity twistor in the definition
of the spinor brackets.
The expansions of the only $\delta$-dependent invariants are
\begin{alignat}{2}
x_{nj}^2&= y_{1 j}^2 + \delta \frac{\l n-1 \ n \ j-1 \ j \r}{\l n-1 \ n \r \l j-1 \ j \r} \quad &\textrm{ for } 2 \leq j \leq n-2 \, ,\\
x_{1j}^2&=y_{1j}^2 +\delta \frac{\l n \ 1 \ j-1 \ j \r}{\l n \ 1 \r\l j-1 \ j \r} &\textrm{ for }  3\leq j \leq n-1 \, ,
\end{alignat}
through subleading order in $\delta$, where the $y_{ij}^2$ are multiparticle invariants with $(n{-}1)$-point kinematics.
\bibliography{bibliography}
\bibliographystyle{utphys}
\end{document}